\documentclass{article}
\usepackage[utf8]{inputenc}
\usepackage{hyperref}
\usepackage{apacite}
\usepackage{comment}
\usepackage{graphicx}
\usepackage{xspace}
\usepackage{float}
\usepackage{fullpage}
\usepackage{amsmath, amssymb}
\usepackage{tcolorbox}
\usepackage{listings}
\usepackage{xcolor}
\usepackage{pifont}
\usepackage{natbib}
\usepackage{xcolor}
\hypersetup{
    colorlinks,
    linkcolor={red!50!black},
    citecolor={blue!50!black},
    urlcolor={blue!80!black}
}

\newcommand{\ImageWidth}{6cm}
\usepackage{tikz}
\usetikzlibrary{calc}
\usetikzlibrary{shadows}
\usetikzlibrary{positioning}
\usetikzlibrary{backgrounds,automata}
\usetikzlibrary{shapes}
\usetikzlibrary{decorations.pathreplacing,positioning, arrows.meta}
\usepackage{subfigure}

\tikzset{fontscale/.style = {font=\relsize{#1}}
    }
    
\definecolor{eclipseStrings}{RGB}{42,0.0,255}
\definecolor{eclipseKeywords}{RGB}{127,0,85}
\colorlet{numb}{magenta!60!black}

\definecolor{lightblack}{gray}{0.95}
\definecolor{lightlight-gray}{gray}{0.92}
\definecolor{darkgray}{gray}{0.80}

\tikzstyle{block} = [rectangle, rounded corners, minimum width=3cm, minimum height=1cm,text centered, draw=black]
\tikzstyle{arrow} = [thick,->,>=stealth]

\lstdefinelanguage{json}{
    basicstyle=\scriptsize\ttfamily,
    commentstyle=\color{eclipseStrings}, 
    stringstyle=\color{eclipseKeywords}, 
    numbers=left,
    numberstyle=\scriptsize,
    stepnumber=1,
    numbersep=7pt,
    showstringspaces=false,
    breaklines=true,
    frame=lines,
    tabsize=3,
    backgroundcolor=\color{white}, 
    string=[s]{"}{"},
    comment=[l]{:\ "},
    morecomment=[l]{:"},
    literate=
        *{0}{{{\color{numb}0}}}{1}
         {1}{{{\color{numb}1}}}{1}
         {2}{{{\color{numb}2}}}{1}
         {3}{{{\color{numb}3}}}{1}
         {4}{{{\color{numb}4}}}{1}
         {5}{{{\color{numb}5}}}{1}
         {6}{{{\color{numb}6}}}{1}
         {7}{{{\color{numb}7}}}{1}
         {8}{{{\color{numb}8}}}{1}
         {9}{{{\color{numb}9}}}{1}
}

\newcommand{\quotes}[1]{``#1''}

\title{Instance generation tool for on-demand transportation problems}
\author{Michell Queiroz {\thanks{Department of Engineering Management, University of Antwerp Operations Research Group (ANT/OR), Prinsstraat 13, 2000, Antwerp,
Belgium.  ({\tt michell.queiroz@uantwerpen.be}). Corresponding author.}} \and Flavien Lucas {\thanks{CERI Numeric Systems, IMT Nord Europe, Institut Mines-Télécom, Univ. Lille, Centre for Digital Systems, F-59000 Lille, France.  ({\tt flavien.lucas@imt-nord-europe.fr})}} \and Kenneth Sörensen {\thanks{Department of Engineering Management, University of Antwerp Operations Research Group (ANT/OR), Prinsstraat 13, 2000, Antwerp,
Belgium.  ({\tt kenneth.sorensen@uantwerpen.be})}} }
\date{\today}

\begin{document}

\maketitle

\begin{abstract}

We present REQreate, a tool to generate instances for on-demand transportation problems. Such problems consist of optimizing the routes of vehicles according to passengers' {demand for transportation} under space and time restrictions (requests). REQreate is flexible and can be configured to generate instances for a large number of problems in this problem class. In this paper, we demonstrate this by generating instances for the Dial-a-Ride Problem (DARP) and On-demand Bus Routing Problem (ODBRP). In most of the literature, researchers either test their algorithms with instances based on artificial networks or perform real-life case studies on instances derived from a specific city or region. Furthermore, locations of requests for on-demand transportation problems are mostly {randomly chosen} according to a uniform distribution. 

The aim of REQreate is to overcome these non-realistic and overfitting shortcomings. Rather than relying on either artificial or limited data, we retrieve real-world street networks from OpenStreetMaps (OSM). To the best of our knowledge, this is the first tool to make use of real-life networks to generate instances for an extensive catalogue of existing and upcoming on-demand transportation problems. Additionally, we present a simple method that can be embedded in the instance generation process to produce distinct urban mobility patterns. We perform an analysis with real life datasets reported by rideshare companies and compare them with properties of synthetic instances generated with REQreate. Another contribution of this work is the introduction of the concept of instance similarity that serves as support to create diverse benchmark sets, in addition to properties (size, dynamism, urgency and geographic dispersion) that could be used to comprehend what affects the performance of algorithms. \\

\noindent \textbf{Keywords:} transportation, instance generator, on-demand public transport, REQreate 
\end{abstract}

\section{Introduction}\label{sec:introduction}

 REQreate is a tool designed with the main objective to generate instances for on-demand transportation problems. Such problems consist of optimizing the routes of vehicles according to passengers' demands under space and time restrictions. Recent years have seen a surge of interest in advanced public transportation systems, and the related planning problems are increasingly being studied in the scientific literature~\citep*{vansteenwegen2022survey}. Many of these problems are NP-hard \citep*{GaJo79}, and the computational effort to optimally solve them often grows exponentially with the size of the instance input. Appropriately, optimization techniques have been proposed to be effective and feasible alternatives for solving these problems.


Instances are the core for evaluating the developed methods for optimization problems. Initially, the instances are essential because they can demonstrate that the method works. Subsequently, they can be used to test how sensitive the approach is to its parameters. Here it is important that the experiments demonstrate that the method is robust and presents good results for different types of instances. Additionally, instances are important to discover the limitations of the approach. A method should be tested both with easy and with difficult instances, and the relationship between the size of the instances and the computation time required to find a good (or even feasible) solution should be established. Lastly, instances are used to compare results with previous approaches, and in some sense, prove that one is better than the other. This is a particular challenge for on-demand public transportation problems which as been shown in a recent survey by \citet*{vansteenwegen2022survey}, who conclude that standard benchmark sets for these problems generally do not exist. This lack of structure complicates any comparison between approaches.  Given these previously mentioned goals, we present REQreate, an instance generation tool {for on-demand transportation problems} based on real-world networks from OpenStreetMaps.  Besides using real-life networks, we will also provide a method to generate requests with different urban mobility properties that result in various scenarios, which the authors believe overcomes the disadvantage of artificially generating data.


 {An instance can be stored as one or multiple files. {Usually}, instances contain a request} database table stored in a Comma Separated Values (CSV) file, where each row represents a passenger's request for transportation and the columns are attributes. An attribute is a piece of information that determines a specific property for each passenger's request. For example, in the case of the Dial-a-Ride Problem (DARP) attributes may include: a) origin location; b) destination location; c) time window for departure; d) time window for arrival; and e) requirement for resources, such as wheelchair or stretcher. Furthermore, the relations between attributes might also constitute a part of the instance, e.g., the travel time/distance matrix between the set of origin and destination locations described in the DARP attributes.

  The tool can assist researchers in testing their optimization algorithms by providing an easy and efficient way to create instances of varying types and sizes. The difficulty in obtaining real-world data is the main reason we decided for a randomly generated approach. There are several reasons to prefer artificially generated instances over real-life data. First, real-life data is difficult to obtain, as companies that implement a specific problem tackled by a researcher are usually not willing to share real cases. The main motives in maintaining such information confidential are to preserve user privacy and gain competitive leverage. One possibility to overcome this issue is to perform large and different surveys, although complexity and the likelihood of the outcome differs from reality are major drawbacks. Second, when an on-demand transportation problem does not yet exist in practice, obtaining real-world data might be impossible. Third, supply for on-demand transportation creates demand, and the real-life data of today might not be representative of tomorrow's situation. Indeed, in \citet*{gkiotsalitis2016demand}, the authors take into account non-recurrent trips, e.g. leisure, which can account for more than half of trips in some cities, to improve the operations of a demand responsive public transportation system.

We developed the tool to be flexible and generic as possible. By providing a configuration file in JavaScript Object Notation (JSON) format, it allows to generate instances for a diversity of problems. Such problems include the Dial-a-Ride Problem (DARP), On-demand Bus Routing Problem (ODBRP), School Bus Routing Problem (SBRP), and many others. Considering that new problems emerge constantly, and absence of test instances is an issue, the tool has the potential to generate instances that could be applied to those novel problems by providing the necessary parameters.


In summary, REQreate was developed with the capability to generate benchmark instance sets that satisfy the following requirements: size, diversity, extensibility, and realism \citep*{vanhoucke2007nsplib}. In this paper, besides presenting the tool, we generate instances and analyze their properties. Two problems are used to demonstrate the potential of REQreate: the DARP and the ODBRP. We discuss how the parameters can be easily tuned to vary the size of instances from small to large, which allows computational studies to derive meaningful conclusions regarding the limitations of the approach, as it is common for larger instances to place a higher computational burden. Moreover, properties of the instances such as size, dynamism, urgency and geographic dispersion are described. 

The diversity of a benchmark set can be evaluated with a measure called instance proximity, similar to an approach introduced in \citet*{leeftink2018case}. Instances are shown to be easily extensible by including or modifying the provided attributes and parameters in the configuration file. Furthermore, the transportation research community is motivated by real-world problems, therefore instances should ideally resemble realistic scenarios. We demonstrate how the synthetic instances compare with real data from rideshare companies trips.

The remainder of this paper is organized as follows. Section~\ref{sec:litreview} presents a literature review on the topic of instance generation. Sections~\ref{sec:retnetwork}~and~\ref{sec:requests} describe the processes to retrieve realistic networks and request generation, respectively. Section~\ref{sec:model} outlines the simple method that can be included in the request generation to produce different urban mobility patterns. Section~\ref{sec:example} introduces formal notations for the DARP and ODBRP used throughout the paper. Section~\ref{sec:properties} describes instance properties: size, dynamism, urgency and geographic dispersion. Section~\ref{sec:similarity} introduces the concept of instance similarity. The analysis between real data and synthetic instances is performed in Section~\ref{sec:comparison}. Ultimately, final remarks are considered in Section~\ref{sec:finalremarks}. 

\section{Literature Review}\label{sec:litreview}


 Throughout the years, tools have been successfully implemented to generate instances for a diversity of problems. \citet*{rardin1993analysis} introduce an instance generator for the Traveling Salesman Problem (TSP). ProGen is a well-known instance generator for precedence and resource-constrained project scheduling problems. The project is described as a network where nodes represent jobs (tasks) and arcs the precedence relations. Jobs may use a set of available, often scarce, resources. ProGen is presented by \citet*{kolisch1995characterization}, and it is based on concepts such as the topology of the network and resource availability to distinguish between easy and hard instances. \citet*{drexl2000progen} introduce an extension of the aforementioned generator primarily aimed to incorporate labor time regulations, making it suitable to generate instances for problems that involve assigning individuals to a number of jobs. Other instance generators have been also proposed for project scheduling problems, such as DANGEN and RanGen \citep*{agrawal1996dagen, demeulemeester2003rangen}. Both generators introduce measures of complexity for the instances. However, RanGen employs a wide range of parameters related to resources and network topology to overcome the shortcomings of previous generators. 

\citet*{cirasella2001asymmetric} introduce new random instance generators for the Asymmetric Traveling Salesman Problem (ATSP). In instances for the ATSP, the distances of moving back and forth between a pair of cities are not always the same. The authors model instances based on real-world applications such as Stacker Crane and Common Superstring Problems. Ultimately, they compare the performance of different algorithms and conclude that there is no dominance over all instance classes. 

\citet*{pellegrini2005instances} present a procedure to generate instances for the Vehicle Routing Problem with Stochastic Demand (VRPSD). Like the Vehicle Routing Problem (VRP), the VRPSD consists of minimize costs while meeting the requirements for delivery of a set of customers. The distinctive characteristic is that before reaching the customer, only a probability distribution of the demand is known. Creation of the instances are supported by databases with information of population and distances between cities from European countries. Scenarios for concentration of customers are inspired by the location of retail stores in European cities, which are proportional to the number of citizens. Demand for each costumer can be generated using uniform and Bernoulli probability distributions. 

Differently from targeting on specific classes of optimization problems, instance generators that focus on providing instances that have particular structural features are referred to as landscape generators. The geometric features of the landscapes are influenced by given parameters. Accordingly, 
\citet*{gallagher2006general} propose a landscape generator for continuous, bound-constrained optimization problems. 
The authors argue that useful conclusions on the performance of heuristic algorithms can be draw by conducting experiments on the landscape space. Other methods for obtaining desired landscapes can be found in the literature \citep*{morrison1999test, michalewicz2000test}, including the generator proposed by \citet*{hernando2015tunable}, which allows to create instances with controlled properties for permutation-based COPs, such as a fixed number of local optima.

\citet*{de2014hydrogen} draw motivation from the absence of high-quality benchmark networks for Water Distribution Network Design (WDND), and propose HydroGen, a tool to generate artificial Water Distribution Networks (WDNs) varying in size and characteristics. The tool provided by the authors allow creating networks with the following distinguished characteristics, which are featured in realistic scenarios: tree-like versus looped network structures; densely populated versus rural areas; domestic versus industrial demand nodes, among others. The artificially generated WDNs with HydroGen are compared with real ones, and according to graph-theoretical indices, they are showed to have a high resemblance. 

A methodology to extend routing instances from the literature to render more realistic scenarios with time-dependant travel times for routing problems is proposed in \citet*{maggioni2014multi}. The authors incorporate real data produced by traffic sensors networks from the city of Turin in Italy, and consider the multi-path TSP with stochastic travel costs to test their technique. 

\citet*{liu2014application} present an instance generator and a memetic algorithm for the Capacitated Arc Routing Problem (CARP). The authors express the need to evaluate the performance of algorithms on classes of instances that resemble realistic scenarios, such as inspection of electric power lines, garbage collection and winter gritting. Their generator controls density, connectedness, degree and distance distribution of the underlying road network. Instances can be further customized by tuning the demand distribution depending on the application. As an example, for the garbage collection problem, the waste amount to be collected can be set as a function of the arc distance combined with population density.

\citet*{macedo2017generator} provides an additional application of instance generators. The authors apply the generator to construct non-regular instances for Semidefinite Programming (SDP). \quotes{SDP refers to convex optimization problems where a linear
function is minimized subject to constraints in the form of linear matrix inequalities}, as defined by \citet*{macedo2017generator}. The authors conduct numerical experiments with the most popular SDP solvers at the time, and discuss the poor efficacy when applied to non-regular instances.   

 \citet{leeftink2018case} develop a novel instance generation procedure for the Surgery Scheduling Problem and point out the lack of widely used benchmark instance sets in healthcare scheduling. Aiming to maximize the diversity of the instances, the authors measure the similarity between two instances over a concept called {instance proximity}. The approach is deterministic and compares instances that are generated based on similar characteristics. Each pair of surgeries between two instances is compared, and they are considered proximate if their expected duration differs less than a given threshold. Ultimately, the authors generate instances based on real-life and theoretical data. The benchmark set is diversified by selecting a subset of instances in which the maximum proximity between them is minimal.
 
 \citet*{ullrich2018generic} attempt to overcome limitations of numerous instance generators described in the literature, which are only suitable to specific problems. The authors propose a versatile tool capable of producing random instances for various discrete optimization problems. They evidence their tool's flexibility by creating instances for the  Traveling Salesman Problem (TSP), Maximum Satisfiability Problem (Max-SAT), and a new load allocation problem based on the Resource-Constrainted Project Scheduling Problem (RCPSP) with time windows. The tool is designed so the generated instances can be easily reproduced by sharing configuration files. Ultimately, the authors aspire that this standardized procedure supports the creation of new instances that are harder and/or larger.  
 
 To the best of our knowledge, this is the first tool that generate instances for an extensive catalogue of existing and upcoming {on-demand transportation} problems. REQreate is also pioneer amongst instance generators to make use of real-life networks. {These networks can be obtained from OpenStreetMaps (OSM), an open-source collaborative mapping project.} OpenStreetMaps is consistently updated and has global coverage, thus is widely used in the transportation literature to perform real-life case studies. For example, \citet*{dingil2018transport} perform an analysis and comparison of transport indicators to conduct an evaluation of different transport strategies used in 151 urban areas. They use open source data from OSM, and present results specially on the correlation of infrastructure and congestion levels. In \citet*{navidi2018comparison}, the authors also use OSM to extract a real-world network used in the simulation study. Results lead to the conclusion of superiority of the demand responsive transit system, such as having the benefit of reducing passengers' perceived travel time. Likewise, \citet*{drakoulis2018gamified} import the network from \quotes{Trikala, Greece} using OSM. The authors study the implementation of an on-demand public bus transportation service on the city.
 
 The instance generation approach described in this paper is the product of attributes, parameters, expressions, and constraints. A similar approach is presented in \citet*{ullrich2018generic}. However, the authors focus more on a generic strategy instead of realism. The tool described in this paper also differs from previous approaches by offering a diverse set of combinations for parameters and attributes that approximately control the output of instance properties. REQreate is available on GitHub\footnote{https://github.com/michellqueiroz-ua/instance-generator}. 


 


\section{Retrieving realistic networks}\label{sec:retnetwork}


In the literature on on-demand public transport problems, researchers frequently test their algorithms with instances based on artificial networks or perform real-life case studies on a specific city or region. For the first case, a shortcoming is that the performance of these algorithms is evaluated in networks with non-realistic patterns. For the second case, it is unlikely to guarantee the robustness of the proposed method, as the results are probably overfitted to the specific case study. Aiming to overcome these obstacles and to provide an easy way to create benchmark instances with realistic networks, we make use of the OSMnx \citep*{boeing2017osmnx} package to retrieve and analyze real-world street networks from OpenStreetMaps (OSM). By simply providing a string, i.e., the name of the area, OSMnx downloads information on the boundaries of the area and creates a graph. Further computations done by REQreate, e.g., distance/travel time matrix between nodes, are done using information obtained with those graphs. On the retrieved network, requests for transportation are subsequently generated.

The street networks built with OSMnx are primal, i.e., nodes are intersections and arcs represent street segments. They are also non-planar, as generally street networks cannot be represented only in two dimensions because of structures such as bridges, tunnels and overpasses. The arcs have weights associated with them, which represent the distance between two nodes. Another important characteristic of the networks is that they are multidigraphs with self-loops, as they are directed and have more than one arc between the same two nodes. An arc that connects a single node to itself is called a self-loop.

\begin{figure}[hhtp]
 \centering
 \subfigure[]{
\includegraphics[scale=0.30] {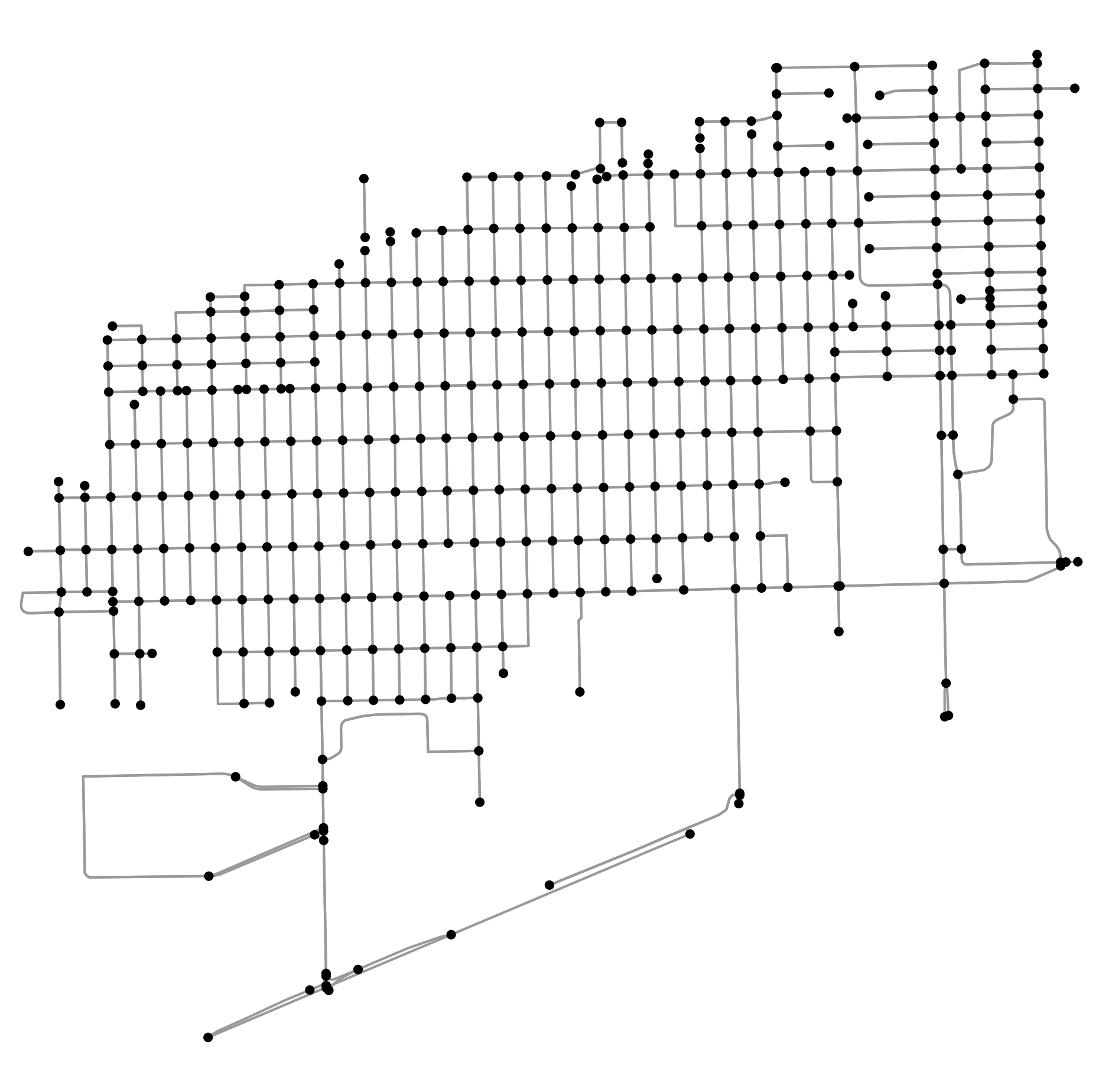}
\label{Fig:networka}

}
 \hspace{0.7cm}
 \subfigure[]{
 \includegraphics[scale=0.30] {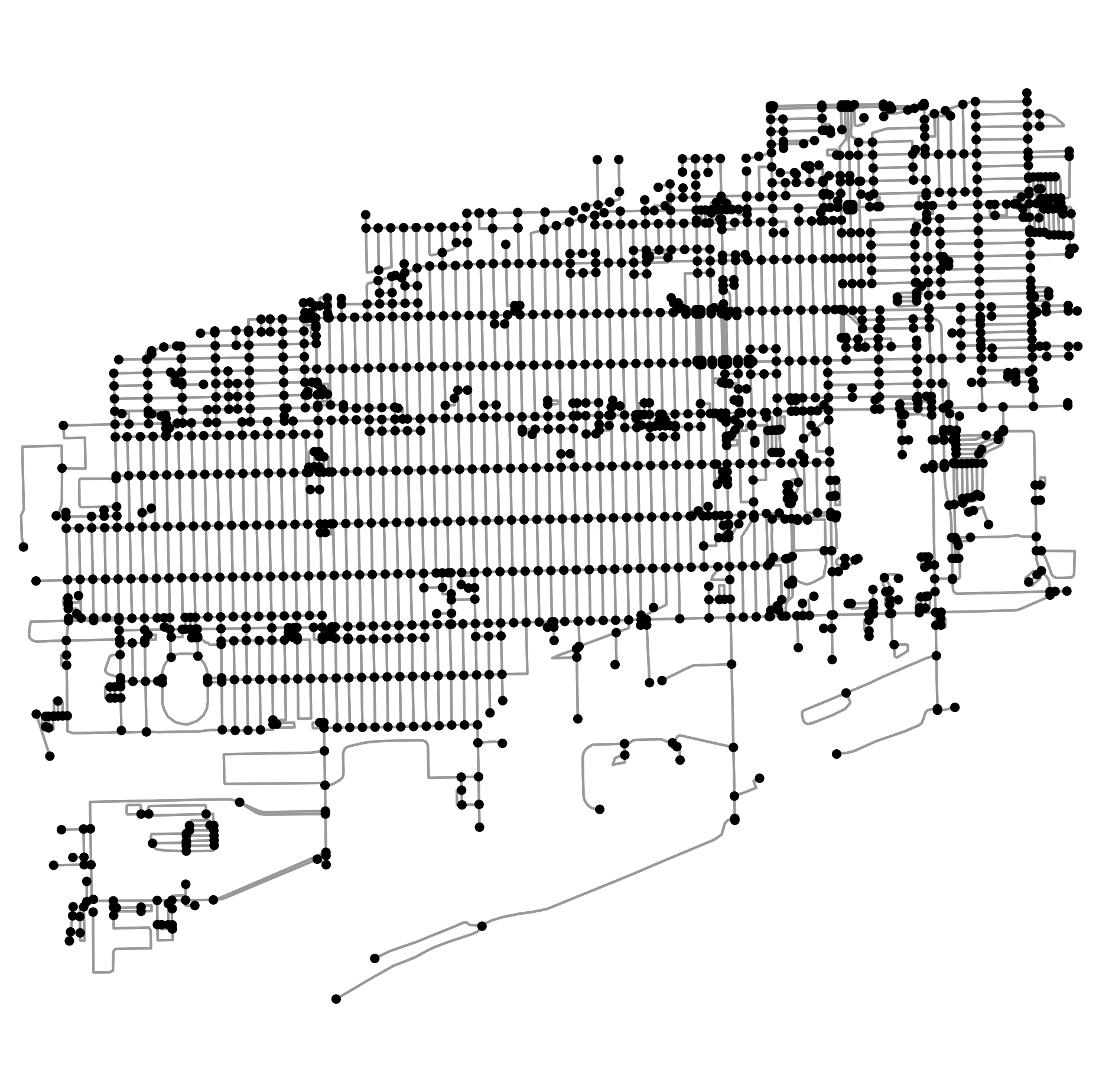}
\label{Fig:networkb}
}
\caption{{Figures~\ref{Fig:networka}~and~\ref{Fig:networkb} depict the driving and walking networks of \quotes{South Lawndale, Chicago, Illinois}, respectively. Note how they are significantly different, as the walking network has more nodes and arcs when compared to the driving one.}}

\label{Fig:network}
\end{figure}

It is possible to download different network types, representing different travel options between nodes. Some examples are: drivable public streets ($drive$), streets and paths that pedestrians can use ($walk$), streets and paths that cyclists can use ($bike$), and others. We work with both drive and walk networks because in reality they have different aspects and distances between nodes. 
An example is shown in Figure~\ref{Fig:network}, where the driving (Figure~\ref{Fig:networka}) and walking (Figure~\ref{Fig:networkb}) networks of a neighborhood (\quotes{South  Lawndale}) in \quotes{Chicago, Illinois} are obtained.

\begin{figure}[hhtp]
 \centering
\includegraphics[scale =0.300] {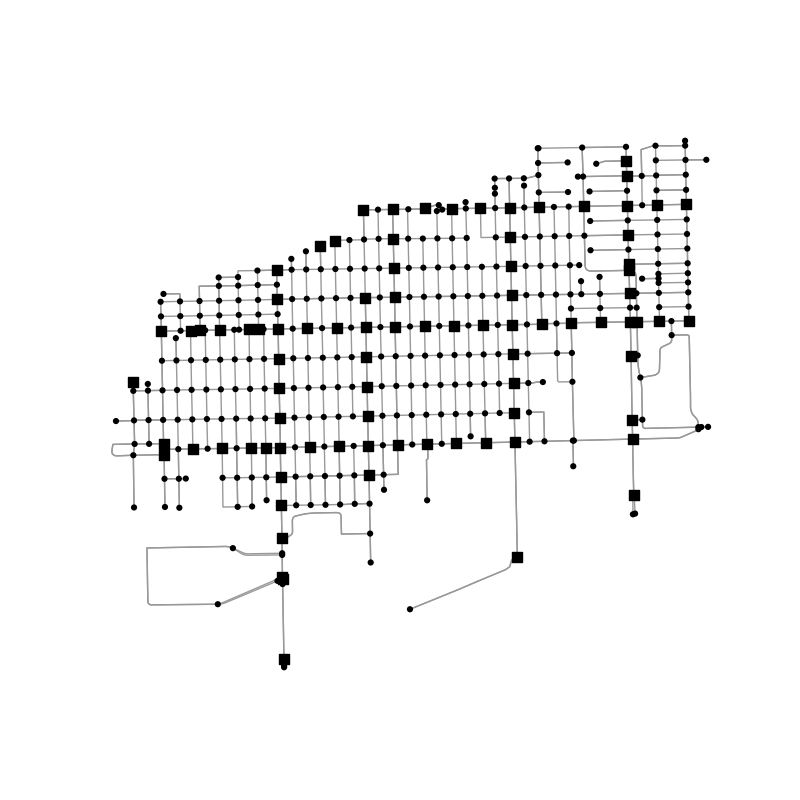}
\caption{{Location of bus stations on the driving network of \quotes{South Lawndale, Chicago, Illinois} represented as black squares.}}
\label{Fig:busstations}
\end{figure}

Bus stations that are within the boundaries of the given area are also obtained using a built in function of OSMnx. This is done because several problems can make use of such information, such as the ODBRP, to assign passengers to pick-up and drop-off stations. After retrieving the bus stations, we perform a verification process to delete repeated entries and stations that are isolated in the network, i.e., unreachable stations. Bus stations on the driving network of \quotes{South  Lawndale, Chicago, Illinois} are represented in Figure~\ref{Fig:busstations}. {Pairing locations between networks is possible with the conversion of geographic coordinates to the nearest node on each respective network. For example, this finds useful when a passenger is assigned to be picked up by a bus, therefore, there is a requirement to determine where approximately the bus should stop (driving network) and the place the passenger needs to wait (walking network).}

{Certain problems may take advantage of the existing fixed route network, which is the case in the Integrated Dial-a-Ride Problem (IDARP) \citep*{hall2009integrated}. However, OSM often does not include fixed line routes, frequency or estimated travel time. Therefore, the collection of data presented in \citet*{kujala2018collection} assists in obtaining such information. The authors published a curated collection of public transport (PT) network data sets for 25 cities, which ultimately provides a testbed for developing tools for PT network studies and PT routing algorithms, and supports an analysis of how PT is organized across the globe.}


The next step is to compute vehicle travel times on the drive network, which requires a speed value. We assume the driving speed to be fixed, which means that the travel times are not dependent on hour of day. 
The NetworkX package \citep*{hagberg2008exploring} allows to access and manage information associated with each node and arc of the network. Let $d_{uv}$ and $ms_{uv}$ be the distance and maximum speed between nodes $u$ and $v$. The travel times are then computed and stored, for each arc in the network, according to the formula: { $tt_{uv} = \alpha \cdot \frac{d_{uv}}{ms_{uv}}$}. Parameter $\alpha$ is called \quotes{speed factor}, is a value in the interval ]0,1] determined by the user, and is used to calculate a realistic speed for each arc, which is usually far from the maximum. The definition of $\alpha$ is demonstrated in Section~\ref{sec:requests}. It is also possible to replace $ms_{uv}$ by an equal speed value to all arcs. 

Travel times are important in order to measure, for example, assuming a door-to-door service, the cost of driving a passenger from their origin to the destination. Another case is to assess the feasibility of assigning passengers to nearby bus stations. Therefore, we build the travel time matrix using Dijkstra's algorithm built in function of the NetworkX package by computing the shortest paths between the nodes of the network using travel times as the weight. {Finally,} OSMNx is also used to download information of school's locations. This data is mostly relevant to generate instances for the SBRP.

\section{Generating requests}\label{sec:requests}

The process of creating instances for the targeted problems in this paper requires generating a set of requests with attributes {randomly chosen} according to probability density functions (pdf) or expressed as relations between each other. Attributes and parameters that define an instance are described in a configuration file given as input to REQreate. Similarly to \citet*{ullrich2018generic}, we also choose the JSON format, considering it is simple, flexible, and easily readable by Python, the language in which our instance generator itself is developed. In this section, we outline the values that should be provided in order to generate requests.
 
 
 \subsection{JSON syntax introduction}
 
JSON files consists of data represented by \textit{name/value} pairs. The field \textit{name} is written in double quotes. The field \textit{value} is separared from \textit{name} by a colon (:), supports primitives data types (strings, numeric, boolean), and more complex structures such as arrays and objects. Arrays are enclosed in square brackets ([~]) and are ordered sets of values. Objects are enclosed in curly brackets (\{~\}) and are collections of \textit{name/value} pairs. The full review of JSON syntax is out of the scope of this paper, therefore we recommend the user to study the examples provided in the tool's GitHub page to acquire familiarity with the format. 
 
 A simple structure of a standard configuration file in JSON format is shown in Listing~\ref{list:skele}. Default data pairs will be referred to as {items}. The names for supported items are: \textit{network}, \textit{seed}, \textit{problem}, \textit{fixed\_lines}, \textit{max\_speed\_factor}, \textit{replicas}, \textit{requests}, \textit{places}, \textit{parameters}, \textit{attributes}, and \textit{instance\_filename}. Pairs enclosed in objects will be referred to as \textit{sub-items}, e.g., in item \textit{parameters}, an object is depicted with 3 sub-items: \textit{name}, \textit{type}, and \textit{value}.

 \begin{lstlisting}[language=json,firstnumber=1, caption={Configuration file structure}, label={list:skele}]
{  
    "network": "Chicago, Illinois",
    "seed": 100,
    ...
    "places": [...],
    "parameters": [ 
        {
            "name": "simple_parameter",
            "type": "integer",
            "value": 10
        },
        {...}
    ],
    "attributes": [...] 
}
  \end{lstlisting}
 
 \subsection{General items}\label{sec:generalitems}
 
  Listing~\ref{list:ex2} illustrates the items \textit{network}, \textit{seed}, \textit{problem}, \textit{fixed\_lines}, \textit{max\_speed\_factor}, \textit{replicas}, \textit{requests}, and \textit{instance\_filename}. First, \textit{network} is a string used to retrieve the network details as described in Section~\ref{sec:retnetwork}. The \textit{seed} consists of a single integer value, and can be used to force the random number generator to generate the same set of numbers at every run of the tool. This allows for reproducibility, meaning that is only necessary to share the configuration files to obtain the same set of instances, instead of the instance files themselves. The problem's acronym can be specified with a string in item \textit{problem}. Information regarding fixed public transport lines can be requested by adjusting the value of binary item \textit{fixed\_lines} to \quotes{true}.  The value of $\alpha$, mentioned in Section~\ref{sec:retnetwork}, can be specified with \textit{max\_speed\_factor}. 
  
  Both \textit{requests} and \textit{replicas} are integer numbers. The former specifies the number of requests in the instance, while {the latter indicates the number of generated request files. Each replica will share the same configuration file, and have e.g., the same number of requests, but the requests themselves will differ because of the randomization process.} 
  The pattern of the instance filename is provided as an array in item \textit{instance\_filename}. The name of any item that has a single primitive as a value can be provided. In Listing~\ref{list:ex2}, the values of \quotes{network}, \quotes{problem}, and \quotes{requests} will constitute the instance filename separated by underscores. Additionally, as each instance must have a unique name, a number from 1 to \textit{replicas} will automatically be printed at the end based on the sequence of generation. Consequently, \quotes{Chicago,Illinois\_DARP\_500\_1} is the filename of the first instance for this particular case.

 \begin{lstlisting}[language=json,firstnumber=1, label={list:ex2}, caption={Example values for items \textit{network}, \textit{seed}, \textit{problem}, \textit{fixed\_lines}, \textit{replicas}, \textit{requests}, and \textit{instance\_filename}}]
{  "network": "Chicago, Illinois",
   "seed": 100,
	"problem": "DARP",
	"fixed_lines": true,
	"max_speed_factor": 0.5,
	"replicas": 10,
	"requests": 500,
	"instance_filename": ["network", "problem", "requests"],
    ...
}
 \end{lstlisting}

 \subsection{Places}
 
The item \textit{places} is an array of objects. This item provides the option to define locations and zones. Specific locations could be used to specify the coordinates of a garage where vehicles are kept waiting to be dispatched (i.e. a depot). Zones are areas within the network and are suitable to characterize a city center, for example. Tables~\ref{table:itemplc} depicts the potential sub-items for \textit{places}. The first column represent the default name. The second column details the data types or possible values of the sub-item. Possible values are differentiated from data types by being written between double quotes. The \textit{name} is the identifier of the place, and could be represented as any sequence of characters between double quotes (string), with the exception of the existing default names. The two options for \textit{type} are \quotes{location} and \quotes{zone}. For a \textit{location}, longitude (\textit{lon}) and latitude (\textit{lat}) must be specified (coordinates in the spherical coordinate system), or \textit{centroid} (center point of the particular network) must be set to \quotes{true}. {Certain locations must have a sub-item called \textit{class}. The only option at the moment for \textit{class} is \quotes{school}, but this can be easily extended. } 

\begin{table}[H]
\centering
\caption{Summary for \textit{places}}
\begin{tabular}{l|l}
\hline
sub-items          & possible values                   \\
\hline
name            & string        \\
type            & \quotes{location}, \quotes{zone} \\
lon       & longitude (geographic coordinate)               \\
lat & latitude (geographic coordinate) \\
centroid & boolean \\
class & \quotes{school} \\
length\_lon  & real \\
length\_lat & real \\
radius & real \\
length\_unit & \quotes{m}, \quotes{km}, \quotes{mi} \\
\hline
\end{tabular}
\label{table:itemplc}
\end{table}

Regarding \textit{zone}, \textit{lon}/\textit{lat} or \textit{centroid} represent the center point of the area. Apart from that, the polygon must also be determined by providing \textit{length\_lon} and \textit{length\_lat} to specify the side lengths of a rectangle/square shape, or \textit{radius} to indicate the radius of a circumference. Finally, \textit{length\_unit} represents the units of length of values declared inside the object, and can be set to \quotes{m}, \quotes{km} or \quotes{mi}, which stands for meters, kilometers and miles, respectively. Listing~\ref{list:ex3} illustrates item \textit{places}. A place is categorized as a \quotes{location} and named \quotes{location\_example}, including \quotes{lon} and \quotes{lat} coordinates equal to -1.6457340723441525 and 48.100199454954804, respectively. Next, a zone has \quotes{zone\_representation} as the identifier, and center coordinates set to -1.6902891077472344 (\quotes{lon}) and 48.09282962664396 (\quotes{lat}). We remark that the tool always verifies if those given coordinates are within the boundaries of the network and raises an error otherwise. The polygon of \quotes{zone\_representation} is established by fixing \textit{length\_lon} and \textit{length\_lat} to 1,000 meters. Lastly, \quotes{zone\_center} is an area stipulated as a circumference with \textit{radius} of 1,500 meters, and \textit{centroid} set to \quotes{true}.


\begin{lstlisting}[language=json,firstnumber=1, label={list:ex3},  caption={Examples for item \textit{places}}]
{   ... 
    "places": [
		{
			"name": "location_example",
			"type": "location",
			"lon": -1.6457340723441525,
			"lat": 48.100199454954804
		},
		{
			"name": "zone_representation",
			"type": "zone",
			"lon": -1.6902891077472344,
			"lat": 48.09282962664396,
			"length_lon": 1000,
			"length_lat": 1000,
			"length_unit": "m" 
		},
		{
			"name": "zone_center",
			"type": "zone",
			"centroid": true,
			"radius": 1500,
			"length_unit": "m"
		},
	],
	...
}
 \end{lstlisting}


 \subsection{Parameters}
 
Item \textit{parameters} is also an array of objects.
Table~\ref{table:itemparam} depicts the potential sub-items for \textit{parameters}. 
The \textit{name} is the identifier of the parameter. Possible options for \textit{type} are \quotes{string}, \quotes{integer}, \quotes{real}, \quotes{array\_primitives}, \quotes{array\_locations} and \quotes{array\_zones}. Sub-item \textit{value} is \textit{type} dependent: a) a sequence of characters for \quotes{string}; b) an integer number for \quotes{integer}; c) a real number for \quotes{real}; d) an array with primitive values for \quotes{array\_primitives}; e) an array with the names of locations declared in \textit{places} for \quotes{array\_locations}; and e) an array with the names of zones declared in \textit{places} for \quotes{array\_zones}. 

\begin{table}[H]
\centering
\caption{Summary for \textit{parameters}}
\begin{tabular}{l|l}
\hline
sub-items          & possible values                   \\
\hline
name            &  string        \\
type            & \quotes{string}, \quotes{integer}, \quotes{real}, \\ & \quotes{array\_primitives}, \quotes{array\_locations},  \quotes{array\_zones}               \\
value       & \textit{type} dependent                  \\
time\_unit & "s", "min", "h" \\
length\_unit & "m", "km", "mi" \\
speed\_unit  & "mps", "kmh", "miph" \\
size & integer \\
locs  & "schools", "random" \\
\hline
\end{tabular}
\label{table:itemparam}
\end{table}

Units of measurement for time, length and speed are expressed in \textit{time\_unit}, \textit{length\_unit}, and \textit{speed\_unit}, respectively. The abbreviations \quotes{s}, \quotes{min} and \quotes{h} for \textit{time\_unit} refer to seconds, minutes and hours, respectively. The possible values for \textit{length\_unit} were previously explained, whereas in \textit{speed\_unit}, \quotes{mps} \quotes{kmh} and \quotes{miph} represent meters per second, kilometers per hour and miles per hour, respectively. We remark the importance of providing these units of measurement, so these values can be properly converted by the generator, since all internal operations are done in meters, seconds, and meters per second.  
Assuming the value for a parameter to be \textit{array\_locations} or \textit{array\_zones}, an integer number must be specified in \textit{size} to delimit its size. In the event of \textit{size} being greater than the number of input names on the given array, random locations will be {randomly chosen} according to \textit{locs}. Values accepted for \textit{locs} are \quotes{random} and \quotes{schools}. We emphasize that in both cases coordinates within the boundaries of the network are {randomly chosen}.

Examples for \textit{parameters} are shown in Listing~\ref{list:ex4}. Parameter \quotes{min\_early\_departure} has \textit{type} established as an integer, its \textit{value} equals to 5, and reported with \quotes{h} (hour) as the unit of time, and in the 24-hour clock format can be interpreted as 5:00. The 24-hour clock format will be the notation used throughout this paper, however in REQreate time is represented in seconds. Representation of a set of locations with \textit{size} 3 can be seen in parameter named \quotes{many\_locations}. The \textit{value} for this parameter is an array containing the name \quotes{location\_example} (defined in Listing~\ref{list:ex3}), and its 2 remaining locations will be randomly chosen. Furthermore, \quotes{pair\_zones} lists 2 zones: \quotes{zone\_representation} and \quotes{zone\_center}, both defined in in Listing~\ref{list:ex3}.

 \begin{lstlisting}[language=json,firstnumber=1, label={list:ex4}, caption={Example for item \textit{parameters}}]
{   ...
	"parameters":[
		{
			"name": "min_early_departure",
			"type": "integer",
			"value": 5,
			"time_unit": "h"
		},
		{
			"name": "many_locations",
			"type": "array_location", 
			"value": ["location_example"],
			"size": 3,
			"locs": "random"
		},
		{
			"name": "pair_zones",
			"type": "array_zones", 
			"value": ["zone_representation", "zone_center"],
			"size": 2
		}
	],
    ...
}
 \end{lstlisting}

 \subsection{Attributes}
 
The \textit{attributes} of an instance are declared in an array of objects. They are usually represented by a range of possible values, or by a relation between other \textit{attributes} and \textit{parameters}. Table~\ref{table:itematt} depicts the potential sub-items for \textit{attributes}. The sub-items \textit{name}, \textit{type}, \textit{time\_unit}, \textit{length\_unit}, and \textit{speed\_unit} are equivalent to the ones previously presented for \textit{places} and \textit{parameters}. 
The sub-item \textit{pdf} (probability density function) is an object to represent a range of values for the attribute, and its structure is detailed in Table~\ref{table:itempdf}. The value of an attribute that carries a \textit{pdf} in its declaration is randomly {chosen} according to the following types of probabilistic distribution functions: \quotes{cauchy}, exponential (\quotes{expon}), \quotes{gamma}, \quotes{gilbrat}, lognormal (\quotes{lognorm}), \quotes{normal}, \quotes{powerlaw}, \quotes{uniform} and \quotes{wald}. Other distributions can be made available in the future. The sub-items \quotes{loc} and \quotes{scale} for \quotes{uniform} indicate the closed interval ([\quotes{loc},\quotes{loc} + \quotes{scale}]) in which values will be {randomly chosen} according to an uniform distribution. Regarding \quotes{normal}, the mean and standard deviation of the distribution are expressed by \quotes{loc} and \quotes{scale}, respectively. 
Besides \quotes{loc} and \quotes{scale}, some functions require an additional parameter \quotes{aux}. For a full rundown on what each parameter means for each distribution, we refer to SciPy's documentation\footnote{https://docs.scipy.org/doc/scipy/index.html}. The declared values must be in conformity with the unit of measurement given for the attribute. Moreover, an attribute can be represented as a mathematical expression transmitted as a string. Restrictions are imposed with \textit{constraints}, which is an array of formulas to be evaluated as true or false. The syntax for \textit{expression} and \textit{constraints} follow Python standards and can contain numbers, identifiers for other attributes or parameters, mathematical symbols, and parentheses. 

\begin{table}[H]
\centering
\caption{Summary for \textit{attributes}}
\begin{tabular}{l|l}
\hline
items          & possible values                   \\
\hline
name            &  string        \\
type            & \quotes{string}, \quotes{integer}, \quotes{real}, \\ & \quotes{location}, \quotes{array\_primitives}               \\
time\_unit & "s", "min", "h" \\
length\_unit & "m", "km", "mi" \\
speed\_unit  & "mps", "kmh", "miph" \\
pdf  & object \\
expression & string \\
constraints  & array of strings \\
subset\_primitives & string \\
subset\_locations  & string \\
subset\_zones  & string \\
weights  & array of numbers \\
output\_csv  & boolean \\
\hline
\end{tabular}
\label{table:itematt}
\end{table}

\begin{table}[H]
\centering
\caption{Summary for \textit{pdf}}
\begin{tabular}{l|l}
\hline
sub-items          & possible values                   \\
\hline
name            &  string        \\
type            &  \quotes{cauchy},  \quotes{expon}, \quotes{gamma}, \\ &         
                     \quotes{gilbrat},  \quotes{lognorm},  \quotes{normal}, \\ & 
                 \quotes{powerlaw}, \quotes{uniform}, \quotes{wald} \\
loc & real \\
scale  & real \\
aux  & real \\
\hline
\end{tabular}
\label{table:itempdf}
\end{table}



Sub-item \textit{subset\_primitives} takes as value an identifier of a parameter previously declared as an array of primitives, and whenever set, the value of the attribute will be chosen considering the given array. Regarding attributes that have \quotes{location} as a type, coordinates will be {randomly chosen} taking into account the full network area. Exceptions are when a method later explained in Section~\ref{sec:model} is set, or when either \textit{subset\_locations} or \textit{subset\_zones} are declared, whose values consist of a string containing the \textit{name} of a parameter stated as an array of locations or zones. Regarding \textit{subset\_locations} a location will be randomly chosen from the declared array, but in the case of \textit{subset\_zones}, one zone is first randomly chosen and then a coordinate within its boundaries is selected. The option to influence the possibility of {randomly choosing} a value from \textit{subset\_primitives}, a location from \textit{subset\_locations}, or a zone from \textit{subset\_zones}, is supported with \textit{weights}, which contains an array of numbers indicating the probability of an element to be selected. Note that the numbers in \textit{weights} must be sequentially declared in accordance with the values of the array it refers to. Item \textit{output\_csv} {indicates} if the attributed is printed in the instance csv file (\quotes{true}, which is default) or not (\quotes{false}). The attributes that are not printed have merely the purpose to assist in the definition of other attributes or to impose more constraints.

Lastly, we provide the opportunity to expresses the interaction of locations through a travel time matrix, i.e., a structure formed by cells indicating the journey time between origin and destination pairs. 
For this purpose, \textit{travel\_time\_matrix} must be reported containing an array with \textit{names} of locations originally declared in \textit{parameters} or \textit{attributes}. The tool computes the trip duration for {every location pair using Dijkstra's algorithm}, and populates the cells of a separate CSV file, whose first row and column are headers indicating the source and target, respectively. A graph is also created, in which nodes are locations, and arcs have an weight associated indicating the travel time. The graph is saved as GraphML format. 

Examples for \textit{attributes} and \textit{travel\_time\_matrix} are depicted in Listing~\ref{list:att}. First, \quotes{earliest\_departure} has type \quotes{integer}, and its values are {randomly chosen according to} a normal probability distribution with mean 30600 seconds (8:30), standard deviation of 3600 seconds (1 hour), and constrained to be greater than or equal to parameter \quotes{min\_early\_departure} (declared in Listing~\ref{list:ex4}). Meanwhile, attribute \quotes{latest\_arrival} is calculated after the mathematical expression \quotes{earliest\_departure + 1800}. The coordinates for \quotes{origin} can be a part of anywhere in the network, whereas for \quotes{destination}, the coordinate will belong to either zone in \quotes{pair\_zones} (see Listing~\ref{list:ex4}), with \quotes{zone\_center} being 3 times more likely to be chosen according to \textit{weights}. The locations that will constitute  \textit{travel\_time\_matrix} are the values of attributes \quotes{origin} and \quotes{destination} of each request.

 \begin{lstlisting}[language=json,firstnumber=1, label={list:att}, caption={Examples for items \textit{attributes} and \textit{travel\_time\_matrix}}]
{   ...
	"attributes": [
		{
			"name": "earliest_departure",
			"type": "integer",
			"time_unit": "s",
			"pdf": {
                    "type": "normal",
                    "loc": 30600,
                    "scale": 3600
				    },
			"constraints": [ "earliest_departure >= min_early_departure"]
		},
		{
			"name": "latest_arrival",
			"type": "integer",
			"time_unit": "s",
			"expression": "earliest_departure + 1800"
		},
		{  
			"name": "origin",
			"type": "location"
		},
		{  
			"name": "destination",
			"type": "location",
			"subset_zones": "pair_zones",
			"weights": [1, 3]
		}
	],
	"travel_time_matrix": ["origin", "destination"]
}
 \end{lstlisting}

\subsection{Invoking method}

{The set of instances will be generated after invoking a method with the configuration file name as an argument} (see Listing~\ref{list:py}). The generator creates one request at a time. Analogous to the technique described in \citet{ullrich2018generic}, a directed graph is built with attributes representing nodes, and arcs defined by expressions or constraints. Then, the attributes (nodes) are topologically sorted in conformity with their dependencies (expressions and constraints). Initially, in accordance with the topological ordering, a value for an attribute is generated. Furthermore, the feasibility is checked based on the disclosed constraints and if any of them is violated, the procedure restarts by discarding the current attribute or the entire request. This process is repeated until all attributes for a request are valid. The instance is then completed upon the given number in \textit{requests} is accomplished. We remark that is beyond the scope of the tool to certify if the set of constraints creates dependencies that can not be met, i.e., every possible combination of attribute values are infeasible. Thus, after a large number of unsuccessful iterations, the generator will halt, raising an error informing it was unable to meet the requirements. 

 \begin{lstlisting}[language=Python, frame=tb, firstnumber=1, label={list:py}, caption={Demonstration of invoking method to generate instances declared in configuration file \quotes{example\_config.json}}]
import input_json

input_json(`example_config.json')
 \end{lstlisting}











\section{Urban mobility patterns}\label{sec:model}

In this section we describe the method to generate various mobility patterns, which can be defined by the different probabilities of a place serving as origin and/or destination of a trip. Previous literature has revealed significant regularity in human mobility patterns, which directly affects process driven by it, such as urban planning, traffic engineering and even epidemic modeling \citep*{song2010limits}. The distribution of distances between positions of two consecutive visited locations (often referred as displacement) have been well approximated by the Levy flight or truncated Levy flight models \citep*{barbosa2018human}.
 
 
 Conventionally, (truncated) Levy flight models are random walks processes in which the step lengths are distributed according to a (truncated) power law tail probability distribution. In addition, travel directions are random and isotropic. More specifically, it has been demonstrated that the step size denoted by $\Delta d$ quantifying the distance between consecutive locations follows a power law distribution $P(\Delta d) \sim \Delta d^{(-1 + \beta)}$ or a truncated power law distribution $P(\Delta d) \sim (\Delta d +  d_{0})^{(-\beta)}exp(-\Delta d/\kappa)$, where  $0 < \beta < 2$, and $d_{0}$ and $\kappa$ symbolizes cutoffs at small and large values of $d$. Given that short trips are more common than long trips, the proven distributions of distance decays are coherent. 
 
 However, Levy flight models have some limitations, as {they do} not account for geographical heterogeneity, which expresses that the probability of a location serving as a possible stop in a journey fluctuates according to geographical space. {In order to overcome these limitations, the conventional Levy flight model can be extended. For example, \citet*{liu2012understanding} represent geographical heterogeneity as a function of population density, which means that high populated areas are expected to attract greater number of trips.}

 {Furthermore, points of interest (POIs) such as airports, parks, shops, offices, among others, can increase the number of trips when compared to the estimation by population density. So, inspired by Stouffer's theory of \textit{intervening opportunities} \citep*{stouffer1940intervening}, which states that \quotes{the number of people going a given distance is directly proportional to the number of opportunities at that distance and inversely proportional to the number of intervening opportunities}, led \citet*{noulas2012tale} to propose the rank-distance model, which was shown to perform well and captured with high accuracy the displacements in an urban environment. }

We integrated ideas from the previous mentioned works in order to create a simple method to approximate urban mobility patterns and generate transportation requests that are more realistic. The purpose is to identify how closely the density of POIs combined with {randomly chosen} distances according to a probabilistic density function can resemble real trips reported by rideshare companies. First, a zone $u$ to contain either the origin/destination (order is uniformly randomly chosen) location is selected with a probability ($P_u$) that is directly proportional to the number of POIs in this zone. We thus have:

\begin{align}
    P_u \propto {number\_POIs(u)}
    \label{form:probognzone}
\end{align}

  Function $number\_POIs(u)$ returns the number of POIs in zone $u$. Then, distances are {randomly chosen} according to the probability distribution function that best fits the data under consideration. To identify the distribution we use the FITTER package~\citep*{fitter}. Finally, {coordinates for a second location} are generated guaranteeing that their distance from the first location approximate the randomly selected distance value, {and at this stage, the direction is randomly chosen according to a uniform distribution.} {In summary, locations are randomly generated considering: a) proportionality to the presence of POIs in the area, which means that regions with greater concentration of POIs will attract more requests; and b) inverse proportionality to the traveled distance, i.e., the higher the distance between two locations lower the probability that they will represent the origin and destination of a request.}
  
  Listing~\ref{list:model} depicts an example of how to integrate the previously mentioned procedure. A supplementary item named \quotes{method\_pois} is included containing two sub-items: \quotes{locations} and \quotes{pdf}. The former contains a list with the two locations from \textit{attributes} that will be either selected according to the higher density of POIs or based on a random distance value from the preceding location, meanwhile the latter describes the parameters of the probabilistic density function to which the distances will be {randomly chosen}. Results of the comparison between real data and synthetic instances are shown later in Section~\ref{sec:comparison}.
  
  
  \begin{lstlisting}[language=json,firstnumber=1, label={list:model}, caption={Demonstration of how the method can be embedded in the instance generation process}]
{   ...
	"attributes": [
        ...
		{  
			"name": "origin",
			"type": "location"
		},
		{  
			"name": "destination",
			"type": "location"
		},
		...
	],
	"method_pois": [
	   {  
		   "locations": ["origin", "destination"],
		   "pdf": {
                    "type": "normal",
                    "loc": 6500,
                    "scale": 5000
			    	 }
		}
	],
    ...
}
 \end{lstlisting}
 

\section{Examples}\label{sec:example}

In this section we describe two problems that the tool can generate instances for. The {on-demand transportation} problems under consideration in Subsections~\ref{subsec:darpinst} and \ref{subsec:odbrpinst} are the DARP and the ODBRP, respectively. We briefly define each problem, and formally present the notations for the attributes that will be used throughout this paper.

\subsection{DARP}\label{subsec:darpinst}

The DARP consists of planning vehicle routes to serve users that indicate origin (pick-up/departure point) and destination (drop-off/arrival point) through requests. The DARP arises in various contexts and has many variations, but one of the most common is the door-to-door transportation service of elderly or disabled people. Different applications yield distinct constraints or objectives, although cost minimization is the most common goal. Variants of the DARP are either static or dynamic. In the static DARP, all request are known beforehand, while in the dynamic case, requests are announced to the system during the day and the vehicle routes must be adjusted in real-time. For complete surveys on research development of DARP variants we refer to \citet*{cordeau2007dial} and \citet*{ho2018survey}. 

{In this paper we consider the dynamic DARP. Instances for this problem} consist of a set of requests $R$ and a travel time matrix. Each request $r \in R$ contains a couple ($o_{r}$, $d_{r}$), which express, respectively, its origin and destination location. The set of origins and destinations of the users are denoted as $O = \bigcup_{r \in R } o_r$ and $D = \bigcup_{r \in R } d_r$, respectively. The moment each request $r$ is announced to the system is denoted time stamp $ts_r$. Additionally, every request {has} two time windows, i.e., the earliest and latest times for pick-up and drop-off. The earliest and latest departure times are denoted $e^u_{r}$ and $l^u_{r}$, respectively. Meanwhile, the earliest and latest arrival times are denoted $e^o_{r}$ and $l^o_{r}$, respectively. Some requests may include requirements for vehicles that support wheelchairs. The fleet of vehicles are located in a set $W$ of designated locations named depots. The travel time matrix consists of the travel times between all pair of locations among the following sets: depot(s) ($W$), origins ($O$) and destinations ($D$). Let $L$ be the complete set of locations, i.e., $L = W \cup O \cup D$. The planning period is expressed as: $T_{e} = [ts_{min}, ts_{max}]$, where $ts_{min}$ and $ts_{max}$ denote, respectively, the earliest time a request can be picked up and latest time it can be dropped off. $T_{e}$ also indicates the period dynamic requests may be announced to the system. An example configuration file for the DARP is showed in Appendix~\ref{app:A}.



\subsection{ODBRP}\label{subsec:odbrpinst}

The ODBRP involves the routing and scheduling of on-demand buses. For this problem, as opposed to being assigned to their origin or destination locations, passengers are picked up and dropped off at designated bus stations. Objectives include to effectively serve all the requests and minimize total user ride time. \citet*{melis2020demand} have recently tackled this problem. The OBDRP can also be operated in the static and dynamic modes.

Instances for the on-demand bus routing problem consist of a set of requests $R$ and a travel time matrix. Each request $r \in R$ figures one passenger and contains a set of potential stops that $r$ can be assigned for pick-up $S^u_r$ and a set of potential stops that $r$ can be assigned for drop-off $S^o_r$. The maximum walking time, representing the passenger's willingness to walk, is denoted by $u_r$. The stations in $S^u_r$ and $S^o_r$ are at most $u_r$ from the origin ($o_{r}$) and destination ($d_{r}$) locations, respectively. Each request also has a time window consisting of an earliest departure time ($e^u_{r}$) and a latest arrival time ($l^o_{r}$). Note that the decision to define a latest departure time ($l^u_r$) is left to the system, therefore the journey is considered feasible as long as it is completed before $l^o_{r}$. $L$ is defined as the complete set of bus station locations. 
The planning period is expressed as: $T_{e} = [ts_{min}, ts_{max}]$, and was previously described in Subsection~\ref{subsec:darpinst}. The travel time matrix consists of the estimated travel times between all the bus stations in $L$. An example configuration file for the ODBRP is showed in Appendix~\ref{app:B}.

\section{Instance properties}\label{sec:properties}

In this section we define several properties of instances for a diversity of {on-demand transportation} problems, including the ones described in Section~\ref{sec:example}. For each instance, we measure the size, value of dynamism, urgency, and geographic dispersion. First, we formally describe these characteristics, and discuss their behavior and impact in realistic scenarios. The terminology used in the following definitions is aligned with the notations of attributes presented in Section~\ref{sec:example}. However, they can be extended to evaluate the properties of other problems that display similar traits. Our main goal is to assist researchers in the investigation of instance characteristics that affect the performance of algorithms, primarily the ones that influence negatively, which eventually supports the development of more robust methods capable of dealing with real life situations. 


\subsection{Size}

  The size of an instance is a measure that depend on to the problem under consideration, e.g., the number of cities determines the size of an instance for the TSP. In previous works addressing the DARP, ODBRP, and similar problems, the instance size is usually expressed as the number of requests \citep*{cordeau2007dial, melis2020demand}. The size of instances is an important aspect, since larger instances frequently impose a greater challenge during experiments. Such behavior is especially evident with exact solvers: as the number of variables increase, the time required to obtain an optimal or even feasible solution grows substantially. Concerning heuristics, the computational time also rises and the solution quality often degrades as instances get larger. Instances of various sizes can be efficiently generated with REQreate, thus used during experiments to evaluate the influence of other variables such as number of available vehicles and their capacity. Testing the limits of a method regarding instance size is crucial, as these systems will be used in real world scenarios where demand can casually increase and the efficiency needs to be maintained.   
\subsection{Dynamism}
\begin{figure}[hhtp]
 \centering
 \subfigure[]{
\begin{tikzpicture}
\draw[thick, -Triangle] (0,0) -- (\ImageWidth,0) node[font=\scriptsize,below left=3pt and -8pt]{time};

\foreach \x in {0,0.5,1,...,5}
\draw (\x cm,6pt) -- (\x cm,-6pt);


\foreach \x in {1,2,3,4,5}
\draw[blue,fill=blue] (\x, 0) circle (1.5pt);

\foreach \x/\descr in {0/0,0.5/1,1/2,1.5/3,2/4,2.5/5,3/6, 3.5/7, 4/8, 4.5/9, 5/10}
\node[font=\scriptsize, text height=2.25ex,
text depth=.5ex] at (\x,-.3) {$\descr$};
\label{Fig:dynamisma}
\end{tikzpicture}
}
 \hspace{0.7cm}
\subfigure[]{

\begin{tikzpicture}
\draw[thick, -Triangle] (0,0) -- (\ImageWidth,0) node[font=\scriptsize,below left=3pt and -8pt]{time};

\foreach \x in {0,0.5,1,...,5}
\draw (\x cm,6pt) -- (\x cm,-6pt);


\foreach \x in {0.5,1,2.5,4,4.5}
\draw[blue,fill=blue] (\x, 0) circle (1.5pt);

\foreach \x/\descr in {0/0,0.5/1,1/2,1.5/3,2/4,2.5/5,3/6, 3.5/7, 4/8, 4.5/9, 5/10}
\node[font=\scriptsize, text height=2.25ex,
text depth=.5ex] at (\x,-.3) {$\descr$};
\label{Fig:dynamismb}
\end{tikzpicture}

}
\hspace{0.7cm}
\subfigure[]{

\begin{tikzpicture}
\draw[thick, -Triangle] (0,0) -- (\ImageWidth,0) node[font=\scriptsize,below left=3pt and -8pt]{time};

\foreach \x in {0,0.5,1,...,5}
\draw (\x cm,6pt) -- (\x cm,-6pt);


\foreach \x in {0.5,1.0,1.5,2.0,2.5}
\draw[blue,fill=blue] (\x, 0) circle (1.5pt);

\foreach \x/\descr in {0/0,0.5/1,1/2,1.5/3,2/4,2.5/5,3/6, 3.5/7, 4/8, 4.5/9, 5/10}
\node[font=\scriptsize, text height=2.25ex,
text depth=.5ex] at (\x,-.3) {$\descr$};
\label{Fig:dynamismc}
\end{tikzpicture}

}
\hspace{0.7cm}
\subfigure[]{

\begin{tikzpicture}
\draw[thick, -Triangle] (0,0) -- (\ImageWidth,0) node[font=\scriptsize,below left=3pt and -8pt]{time};

\foreach \x in {0,0.5,1,...,5}
\draw (\x cm,6pt) -- (\x cm,-6pt);


\foreach \x in {0.5}
\draw[blue,fill=blue] (\x, 0.0) circle (1.5pt);

\foreach \x in {0.5}
\draw[blue,fill=blue] (\x, 0.15) circle (1.5pt);

\foreach \x in {0.5}
\draw[blue,fill=blue] (\x, 0.30) circle (1.5pt);

\foreach \x in {0.5}
\draw[blue,fill=blue] (\x, 0.45) circle (1.5pt);

\foreach \x in {0.5}
\draw[blue,fill=blue] (\x, 0.60) circle (1.5pt);

\foreach \x/\descr in {0/0,0.5/1,1/2,1.5/3,2/4,2.5/5,3/6, 3.5/7, 4/8, 4.5/9, 5/10}
\node[font=\scriptsize, text height=2.25ex,
text depth=.5ex] at (\x,-.3) {$\descr$};
\label{Fig:dynamismd}
\end{tikzpicture}
}
\caption{Depiction of different levels of dynamism. Colored dots represent the time one request was announced to the system. The scenario in Figure~\ref{Fig:dynamisma} changes {continuously in evenly timed intervals}, therefore it represents a very dynamic scenario. The dynamism levels decreases in Figure~\ref{Fig:dynamismb}~and~~\ref{Fig:dynamismc}, {as both scenarios have larger intervals without new announcements.  In Figure~\ref{Fig:dynamismd} all information is known at the same time which results in a case with no dynamism.}}

\label{Fig:dynamism}
\end{figure}
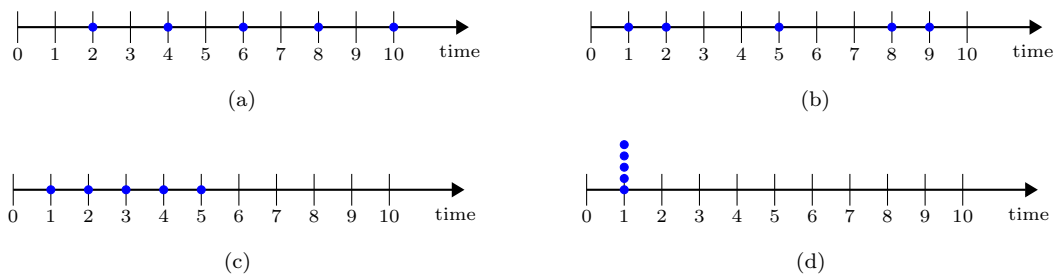

Dynamism captures the periodicity aspect of new information revealed during the planning period. Regarding the problems contemplated in this paper, this new information comes in the form of transportation requests issued by users of the system. Therefore, dynamism is measured as the frequency that new requests are revealed to the system. In a highly dynamic scenario requests will arrive continuously. On the contrary, {if there are long intervals without new information}, the instance can be considered less dynamic. Different scenarios with various levels of dynamism are represented in Figure~\ref{Fig:dynamism}. 


According to observations made by \citet*{kilby1998dynamic} and \citet*{pillac2013review}, direct implications of dynamism are the required number of restarts and the available time for optimization, i.e., constant announcements imply limited time to perform iterations intended to improve the solution. \citet*{borndorfer1999telebus} comment on how planning operations for the DARP are impacted by dynamism. The authors remark that the schedule executed during operational hours is often different from the computed version for the static problem, because of several circumstances, such as new requests, cancellations, accidents, and many others. So, since static problems are usually unrealistic, dynamism and its consequences should be investigated.  

 \citet*{lund1996vehicle} presented an earlier definition where dynamism is the proportion between dynamic and static requests. Since the relative timing distribution of arrivals are not accounted in the measure, scenarios where no previous information is known before the planning period can not be distinguished.  In  \citet*{larsen2002partially}, a request is reportedly more dynamic when revealed to the system with a tight interval between the time stamp and the latest possible time to begin servicing the request. However, as discussed later, the authors actually incorporate the urgency feature in their definition, leading to difficulties in drawing separate conclusions on the correlation of dynamism and urgency related to solution quality. More recently, \citet*{van2016measures} provided separate measures for dynamism and urgency that addresses these limitations, which will be presented in the following paragraphs. 

In order to present the dynamism measure, we start by giving some supporting definitions. Consider $R_{d} = \{r_{1}, r_{2}, ..., r_{|R_{d}|} \}$ to be the set of requests introduced after the start of planning period (dynamic requests), and sorted by non decreasing order of time stamps, i.e., $ts_{r_{j}} \geq ts_{r_{i}}, \ \forall j > i$. Let $\Delta = \{ \delta_{1}, \delta_{2}, ..., \delta_{|R_{d}| - 1} \} = \{ ts_{r_{j}} - ts_{r_{i}} | \ j = i + 1 \  \forall r_{i}, r_{j} \in R_{d} \}$ be the set of interarrival times. Cases with 100\% dynamism are characterized by requests being presented in evenly timed intervals (Figure~\ref{Fig:dynamisma}), thus possessing the perfect interarrival time ($\theta$), computed as follows: $\theta = \frac{T_{e}}{|R_{d}|}$. For each interarrival $\delta_{k} \in \Delta$ is now possible to compute its deviation ($\sigma_{k}$) from the 100\% dynamism case: 

\begin{align}
\sigma_{k} = \begin{cases}
\theta - \delta_{k} & \text{if $(k = 1)$ $\wedge$ $(\delta_{k} < \theta)$} \\
\theta - \delta_{k} + \frac{\theta - \delta_{k}}{\theta} \cdot \sigma_{k - 1} & \text{if $(k > 1)$ $\wedge$ $(\delta_{k} < \theta)$} \\
0 & \text{otherwise}
\end{cases}
\end{align}

The deviation of an entire scenario ($\lambda$) is then given by the following summation: $\lambda = \sum_{\delta_{k} \in \Delta} \sigma_{k}$. Consider bursts as announcements that occur in short periods, consequently leading to iterarrival times smaller than $\theta$. Note that the therm $\frac{\theta - \delta_{k}}{\theta} \cdot \sigma_{k - 1}$ penalizes those bursts by adding a proportion of the deviation from the previous interarrival time. Moreover,
consider $\eta = \sum_{\delta_{k} \in \Delta} \overline{\sigma}_{k}$, where:

\begin{align}
\overline{\sigma}_{k} = \theta +  \begin{cases}
\frac{\theta - \delta_{k}}{\theta} \cdot \sigma_{k - 1} & \text{if $(k > 1)$ $\wedge$ $(\delta_{k} < \theta)$} \\
0 & \text{otherwise}
\end{cases}
\end{align}

The factor $\eta$ theoretically captures the maximum deviation for a scenario (0\% dynamism), and plays the role of normalizing the deviation from the 100\% case. Therefore, dynamism of an instance is measured by:

\begin{align}
    \rho = 1 - \frac{\lambda}{\eta}
    \label{form:dynamism}
\end{align}

{Following the provided definitions, we now can compute the dynamism for scenarios shown in Figure~\ref{Fig:dynamism}. Each of these scenarios exhibit 5 dynamic requests and a perfect interarrival time of 2, i.e., $\theta = \frac{10.00}{5.00} = 2.00$. The latter permits us to compute the necessary sets and values to determine the dynamism, which are reported in  Table~\ref{table:dynamism}. In this table, the first column specifies the scenario (Figures~\ref{Fig:dynamisma}-\ref{Fig:dynamismd}). The second and third columns indicate the set of deviation values ($\sigma$) and its summation ($\lambda$), respectively. Meanwhile, fourth and fifth column report the normalization set values ($\overline{\sigma}$) and its summation ($\eta$), respectively. The last column in the table gives the dynamism value ($\rho$) for the corresponding scenario. Note that Figure~\ref{Fig:dynamisma} represents the 100\% dynamism case, since all interarrival times are equal to $\theta$. Furthermore, dynamism levels decrease in Figures~\ref{Fig:dynamismb}~and~\ref{Fig:dynamismc}, as they exhibit requests that are revealed close to one another ($\delta_{k} < \theta$). Ultimately, the 5 requests depicted in Figure~\ref{Fig:dynamismd} were announced at the same time, so in such circumstance, the dynamism is zero.}

\begin{table}[H]
\small
\centering
\caption{Sets and values necessary to determine dynamism for scenarios shown in Figure~\ref{Fig:dynamism}.}
\begin{tabular}{|l|c|c|c|c|c|c|}
\hline
Scenario                                             & $\Delta$                                & $\sigma$                               & $\lambda$              & $\overline{\sigma}$                                                      & $\eta$                                 & $\rho$                \\
\hline
Figure~\ref{Fig:dynamisma} & $\{ 2.00, 2.00, 2.00, 2.00 \}$ & $\{0.00, 0.00, 0.00, 0.00\} $ & $ 0.00$       & $ \{ 2.00, 2.00, 2.00, 2.00 \}$                       & $ 8.00$                          & $  1.00$        \\

Figure~\ref{Fig:dynamismb} & $\{ 1.00, 3.00, 3.00, 1.00 \}$ & $\{1.00, 0.00, 0.00, 1.00\} $ & $2.00$       & $ \{ 2.00, 2.00, 2.00, 2.00 \}$                       & $ 8.00$                          & $0.75$         \\

Figure~\ref{Fig:dynamismc} & $\{ 1.00, 1.00, 1.00, 1.00 \}$ & $ \{1.00, 1.50, 1.75, 1.87\} $ & $6.12$ & $\{ 2.00, 2.50, 2.75, 2.87 \}$                       & $ 10.12$                   & $0.39$ \\
Figure~\ref{Fig:dynamismd} & $\{ 0.00, 0.00, 0.00, 0.00 \}$ & $\{2.00, 4.00, 6.00, 8.00\} $ & $20.00$      & $ \{2.00, 4.00, 6.00, 8.00\}$ & $20.00$ & $0.00$    \\
\hline
\end{tabular}
\label{table:dynamism}
\end{table}

\subsection{Urgency}
\begin{figure}[hhtp]
 \centering
 \subfigure[]{
\begin{tikzpicture}
\draw[thick, -Triangle] (0,0) -- (\ImageWidth,0) node[font=\scriptsize,below left=3pt and -8pt]{time};

\foreach \x in {0,1,...,4}
\draw (\x cm,6pt) -- (\x cm,-6pt);

\foreach \y/\descr in {1/ts_{r_{i}},4/l^u_{r_{i}}}
\node[font=\scriptsize, text height=-2.25ex,
text depth=.5ex] at (\y,.3) {$\descr$};

\foreach \x/\descr in {0/0,1/1,2/2,3/3,4/4}
\node[font=\scriptsize, text height=2.25ex,
text depth=.5ex] at (\x,-.3) {$\descr$};

\draw [thick ,decorate,decoration={brace,amplitude=5pt}] (1,0.7)  -- +(3,0) 
       node [black,midway,above=4pt, font=\scriptsize] {$f_{r_{i}} = 3$};
\label{Fig:urgencya}
\end{tikzpicture}
}
 \hspace{0.7cm}
\subfigure[]{

\begin{tikzpicture}
\draw[thick, -Triangle] (0,0) -- (\ImageWidth,0) node[font=\scriptsize,below left=3pt and -8pt]{time};

\foreach \x in {0,1,...,4}
\draw (\x cm,6pt) -- (\x cm,-6pt);

\foreach \y/\descr in {1/ts_{r_{j}},2/l^u_{r_{j}}}
\node[font=\scriptsize, text height=-2.25ex,
text depth=.5ex] at (\y,.3) {$\descr$};

\foreach \x/\descr in {0/0,1/1,2/2,3/3,4/4}
\node[font=\scriptsize, text height=2.25ex,
text depth=.5ex] at (\x,-.3) {$\descr$};

\draw [thick ,decorate,decoration={brace,amplitude=5pt}] (1,0.7)  -- +(1,0) 
       node [black,midway,above=4pt, font=\scriptsize] {$f_{r_{j}} = 1$};
\label{Fig:urgencyb}
\end{tikzpicture}

}
\caption{Depiction of different urgency scenarios. Figures~\ref{Fig:urgencya}~and~\ref{Fig:urgencyb} exhibits the reaction times for requests $r_{i}$ and $r_{j}$, respectively. Note that $f_{r_{j}} < f_{r_{i}}$, therefore $r_{j}$ is more urgent than $r_{i}$.}

\label{Fig:urgency}
\end{figure}
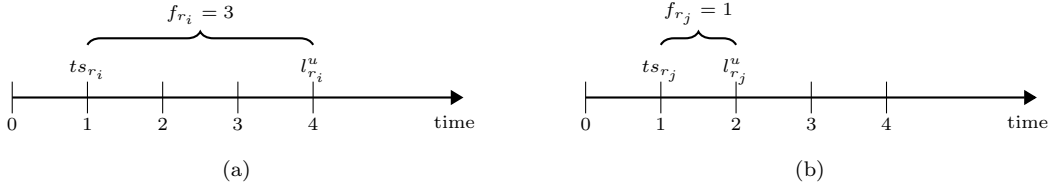

The interval between the time stamp of a request and the latest pickup time is referred as reaction time. Urgency is a feature that indicates the length of this interval, i.e., the available time to perform actions regarding a new dynamic request. Let $\chi = \{f_{r_{1}}, f_{r_{2}}, ..., f_{r_{|R_{d}|}}\}$ be the set of urgency values of an instance, where the urgency ($f_{r_{i}}$) of a single request $r_{i}$ is computed as follows: $f_{r_{i}} = l^u_{r_{i}} - ts_{r_{i}}$. Consider $f_{r_{i}}$ and $f_{r_{j}}$ as the urgency of requests $r_{i}$ and $r_{j}$, respectively. Supposing $f_{r_{i}} < f_{r_{j}}$, $r_{i}$ is said to be more urgent than $r_{j}$. Accordingly, $r_{j}$ is considered to be less urgent. The mean ($\overline{\chi}$) {and standard deviation ($\chi_{s}$)} of $\chi$ denote the urgency of the corresponding instance. They are computed as follows: {$\overline{\chi} = \frac{\sum_{r_{i} \in R_{d}} f_{r_{i}}}{|\chi|}$ and $\chi_{{s}} = \sqrt{\frac{\sum_{r_{i} \in R_{d}} (f_{r_{i}} - \overline{\chi})^2}{|\chi|}}$.} {The standard deviation is part of the urgency measure because the mean alone is not a very distinctive feature, since scenarios with the same mean and different standard deviations may contain different numbers of higher and lower urgency values.} Note that urgency is expressed in time units. The concept of urgency is depicted in Figure~\ref{Fig:urgency}, {which represents a scenario with the following set of urgency values: $\chi = \{3, 1\}$. As a result, the urgency of this instance is: $\overline{\chi} = \frac{3 + 1}{2} = 2$ and  $\chi_{{s}} = \sqrt{\frac{1 + 1}{2}} = 1$.}

\textcolor{blue}{}

\citet{van2016measures} perform computational experiments to evaluate the impact of dynamism and urgency on route quality for the dynamic Pickup and Delivery Problem with Time Windows (PDPTW). In the PDPTW, a fleet of vehicles is used to transport items from pickup to delivery locations according to set of requests made by costumers. The objective is to minimize route costs, which is inversely proportional to route quality. 
The instances generated by the authors were grouped by sets that share similar characteristics, apart from values of dynamism and urgency. The investigated hypotheses led to conclude that these features have different influences on operating costs, as dynamism has a reasonably small effect on operating costs but is negatively correlated with it, while urgency is positively correlated. 

\subsection{Geographic dispersion}
Geographic dispersion is a criterion that express the spread of important locations across the network. It can be expected that a large geographic dispersion, and thus a large distance between important locations in the network, will give rise to longer routes. The definition presented in this paper is based on the approach described in \citet*{reyes2018meal}, although we perform slight modifications to make it suitable for the DARP and ODBRP. First, we sum the estimated direct travel times for each request and calculate their average: $\mu = \frac{\sum_{r_{i} \in R_{}} {tt_{o_{r_{i}}d_{r_{i}}}}}{|R|}$, {where for the DARP $tt_{o_{r_{i}}d_{r_{i}}}$ represents the estimated direct travel time between origin and destination, while for the ODBRP it depicts an estimated average travel time between stations surrounding the origin and destination.} {Since, requests can be served simultaneously by the vehicles, we incorporate the average travel time from the origin and destination to the nearest neighbors locations of requests, namely $\omega$. First, consider $L^o_{r_{i}}$ and $L^d_{r_{i}}$ to be subset of locations that can potentially be served after the origin and destination of $r_{i}$, respectively. Consider $th_{s}$ to be a time threshold indicating that the time windows are close enough to possibly coincide. Specifically, $L^o_{r_{i}} = \{ o_{r_{j}} \ | \ \lVert e^u_{r_{i}} - e^u_{r_{j}} \lVert < th_{s} \ \forall r_{j} \in R \setminus r_{i} \} \cup \{ d_{r_{j}} | \ \lVert e^u_{r_{i}} - l^o_{r_{j}} \lVert < th_{s} \  \forall r_{j} \in R \setminus r_{i} \} $ and $L^d_{r_{i}} = \{ o_{r_{j}} \ | \ \lVert l^o_{r_{i}} - e^u_{r_{j}} \lVert < th_{s} \ \forall r_{j} \in R \setminus r_{i} \} \cup \{ d_{r_{j}} | \ \lVert l^o_{r_{i}} - l^o_{r_{j}} \lVert < th_{s} \  \forall r_{j} \in R \setminus r_{i} \} $. The at most $n$ nearest locations in $L^o_{r_{i}}$ and $L^d_{r_{i}}$ are depicted as $N^o_{r_{i}}$ and $N^d_{r_{i}}$, respectively. Let $tn^{o}_{r_{i}}$ and $tn^{d}_{r_{i}}$ correspond to the average travel times from the origin and destination of $r_{i}$ to locations in $N^o_{r_{i}}$ and $N^d_{r_{i}}$:}

\begin{align}
tn^{o}_{r_{i}} = \begin{cases}
\frac{\sum_{v \in N^o_{r_{i}}} tt_{o_{r_i}v}}{|N^o_{r_{i}}|}, & \text{if $|N^o_{r_{i}}| > 0$} \\
0, & \text{otherwise}
\label{eq:4}
\end{cases} \\
tn^{d}_{r_{i}} =\begin{cases}
\frac{\sum_{v \in N^d_{r_{i}}} tt_{d_{r_i}v}}{|N^d_{r_{i}}|}, & \text{if $|N^d_{r_{i}}| > 0$} \\
0, & \text{otherwise}
\label{eq:5}
\end{cases}
\end{align}

Accordingly, $\omega$ is used to approximate the duration of detours and can be computed as follows: $\omega = \frac{\sum_{r_{i} \in R} {tn^{o}_{r_{i}} + tn^{d}_{r_{i}}}}{2 \cdot |R|}$. 
The definition of geographic dispersion ($gd$) then is calculated as:

\begin{align}
   {gd = \mu + \omega}
\end{align}

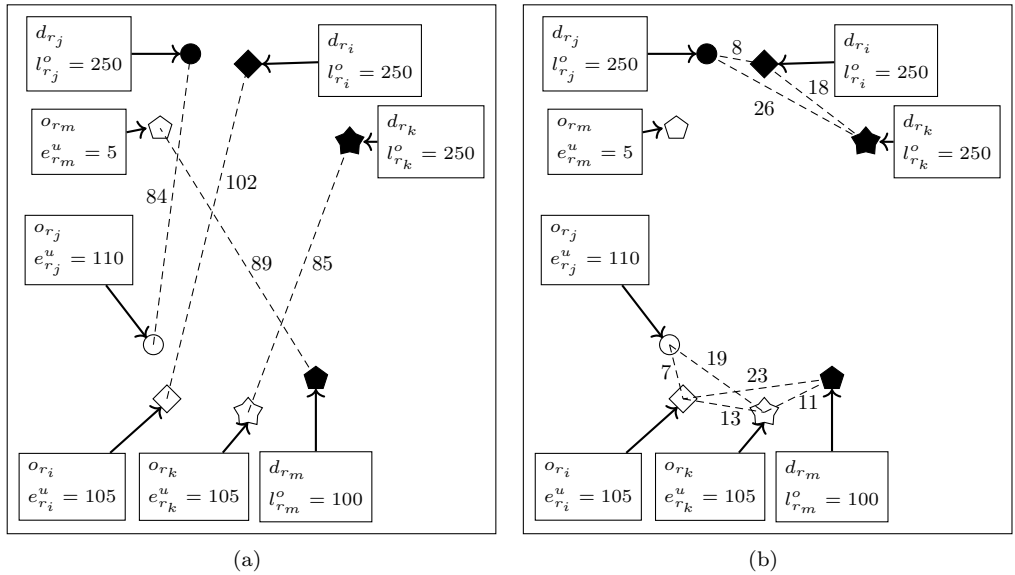
\begin{figure}[hhtp]
\centering
 \subfigure[]{

    \begin{tikzpicture}[framed, scale=0.90]
    \node at (1.6, -2.4) [circle, draw, scale=0.8] {};
    \node at (2.15, 1.9) [circle, draw, scale=0.8, fill=black] {};
    \draw[densely dashed, -] (1.6,-2.4) -- (2.15, 1.9);
    \node at (1.65, -0.2) [scale = 0.8] {84};
    
    \node at (1.8, -3.2) [diamond, draw, scale=0.8] {};
    \node at (3, 1.75) [diamond, draw, scale=0.8, fill=black] {};
     \draw[densely dashed, -] (1.8, -3.2) -- (3, 1.75);
      \node at (2.9, 0) [scale = 0.8] {102};
    
    \node at (3, -3.4)  [star, draw] [scale = 0.8] {};
     \node at (4.5, 0.6)  [star, draw, fill=black] [scale = 0.8] {};
    \draw[densely dashed, -] (3, -3.4) -- (4.5, 0.6);
    \node at (4.1, -1.2) [scale = 0.8] {85};
    
    \node at (4, -2.9)  [regular polygon, draw, fill=black] [scale = 0.8] {};
     \node at (1.7, 0.8)  [regular polygon, draw] [scale = 0.8] {};
      \draw[densely dashed, -] (4, -2.9) -- (1.7, 0.8);
    \node at (3.2, -1.2) [scale = 0.8] {89};
    
     \node[rectangle, draw, align=left] (arr1) at (0.5,-1.0) {\scriptsize 
     $o_{r_{j}}$ \\
      \scriptsize $e^u_{r_{j}} = 110$
     };
     \draw[thick,->] (arr1) -- (1.5,-2.3);
      \node[rectangle, draw, align=left] (arr2) at (0.4,-4.5) {\scriptsize 
     $o_{r_{i}}$ \\
      \scriptsize $e^u_{r_{i}} = 105$
     };
     \draw[thick,->] (arr2) -- (1.72,-3.32);
      \node[rectangle, draw, align=left] (arr3) at (2.2,-4.5) {\scriptsize 
     $o_{r_{k}}$ \\
      \scriptsize $e^u_{r_{k}} = 105$
     };
     \draw[thick,->] (arr3) -- (3,-3.55);
      \node[rectangle, draw, align=left] (arr4) at (4.0,-4.5) {\scriptsize 
     $d_{r_{m}}$ \\
      \scriptsize $l^o_{r_{m}} = 100$
     };
     \draw[thick,->] (arr4) -- (4,-3.05);
     \node[rectangle, draw, align=left] (des1) at (0.5,1.9) {\scriptsize 
     $d_{r_{j}}$ \\
      \scriptsize $l^o_{r_{j}} = 250$
     };
     \draw[thick,->] (des1) -- (2.00,1.9);
     \node[rectangle, draw, align=left] (des4) at (0.5,0.6) {\scriptsize 
     $o_{r_{m}}$ \\
      \scriptsize $e^u_{r_{m}} = 5$
     };
     \draw[thick,->] (des4) -- (1.5,0.8);
     \node[rectangle, draw, align=left] (des2) at (4.8,1.8) {\scriptsize 
     $d_{r_{i}}$ \\
      \scriptsize $l^o_{r_{i}} = 250$
     };
     \draw[thick,->] (des2) -- (3.2,1.75);
     \node[rectangle, draw, align=left] (des3) at (5.7,0.6) {\scriptsize 
     $d_{r_{k}}$ \\
      \scriptsize $l^o_{r_{k}} = 250$
     };
     \draw[thick,->] (des3) -- (4.7,0.6);
\end{tikzpicture}
\label{fig:gda}
}
 \subfigure[]{

    \begin{tikzpicture}[framed, scale=0.90]
    \node at (1.6, -2.4) [circle, draw, scale=0.8] {};
    \node at (2.15, 1.9) [circle, draw, scale=0.8, fill=black] {};
    
    \draw[densely dashed, -] (1.6,-2.4) -- (1.8, -3.2);
      \draw[densely dashed, -] (2.15,1.9) -- (3, 1.75);
    \node at (3.8, 1.4) [scale = 0.8] {18};
    \draw[densely dashed, -] (1.6, -2.4) -- (3, -3.4);
      \node at (2.3, -2.6) [scale = 0.8] {19};
    
    \draw[densely dashed, -] (2.15, 1.9) -- (4.5, 0.6);
      \node at (3, 1.1) [scale = 0.8] {26};
      
    \node at (1.8, -3.2) [diamond, draw, scale=0.8] {};
    \node at (3, 1.75) [diamond, draw, scale=0.8, fill=black] {};
     \draw[densely dashed, -] (1.8, -3.2) -- (3, -3.4);
     \draw[densely dashed, -] (3, 1.75) -- (4.5, 0.6);
      \node at (2.9, -2.85) [scale = 0.8] {23};
      \node at (2.6, 2) [scale = 0.8] {8};
      \node at (2.5, -3.48) [scale = 0.8] {13};
      \node at (1.55, -2.8) [scale = 0.8] {7};
      \node at (3.65, -3.25) [scale = 0.8] {11};

    \node at (3, -3.4)  [star, draw] [scale = 0.8] {};
     \node at (4.5, 0.6)  [star, draw, fill=black] [scale = 0.8] {};
    \draw[densely dashed, -] (3, -3.4) -- (4, -2.9);

    \node at (4, -2.9)  [regular polygon, draw, fill=black] [scale = 0.8] {};
     \node at (1.7, 0.8)  [regular polygon, draw] [scale = 0.8] {};
      \draw[densely dashed, -] (4, -2.9) -- (1.8, -3.2);

     \node[rectangle, draw, align=left] (arr1) at (0.5,-1.0) {\scriptsize 
     $o_{r_{j}}$ \\
      \scriptsize $e^u_{r_{j}} = 110$
     };
     \draw[thick,->] (arr1) -- (1.5,-2.3);
      \node[rectangle, draw, align=left] (arr2) at (0.4,-4.5) {\scriptsize 
     $o_{r_{i}}$ \\
      \scriptsize $e^u_{r_{i}} = 105$
     };
     \draw[thick,->] (arr2) -- (1.72,-3.32);
      \node[rectangle, draw, align=left] (arr3) at (2.2,-4.5) {\scriptsize 
     $o_{r_{k}}$ \\
      \scriptsize $e^u_{r_{k}} = 105$
     };
     \draw[thick,->] (arr3) -- (3,-3.55);
      \node[rectangle, draw, align=left] (arr4) at (4.0,-4.5) {\scriptsize 
     $d_{r_{m}}$ \\
      \scriptsize $l^o_{r_{m}} = 100$
     };
     \draw[thick,->] (arr4) -- (4,-3.05);
     \node[rectangle, draw, align=left] (des1) at (0.5,1.9) {\scriptsize 
     $d_{r_{j}}$ \\
      \scriptsize $l^o_{r_{j}} = 250$
     };
     \draw[thick,->] (des1) -- (2.00,1.9);
     \node[rectangle, draw, align=left] (des4) at (0.5,0.6) {\scriptsize 
     $o_{r_{m}}$ \\
      \scriptsize $e^u_{r_{m}} = 5$
     };
     \draw[thick,->] (des4) -- (1.5,0.8);
     \node[rectangle, draw, align=left] (des2) at (4.8,1.8) {\scriptsize 
     $d_{r_{i}}$ \\
      \scriptsize $l^o_{r_{i}} = 250$
     };
     \draw[thick,->] (des2) -- (3.2,1.75);
     \node[rectangle, draw, align=left] (des3) at (5.7,0.6) {\scriptsize 
     $d_{r_{k}}$ \\
      \scriptsize $l^o_{r_{k}} = 250$
     };
     \draw[thick,->] (des3) -- (4.7,0.6);
\end{tikzpicture}
\label{fig:gdb}
}
\caption{Illustration of an instance with 4 requests to support computations for geographic dispersion.}
\label{fig:gd}
\end{figure}

We exemplify geographic dispersion of an instance using Figure~\ref{fig:gd}. The instance consists of 4 requests $R = \{ r_{i}, r_{j}, r_{k}, r_{m} \}$. Assume $th_{s} = 10$ and $n = 2$. Travel times between locations are represented by the numbers adjacent to the dashed lines. These values are measured in fictitious units of time. First, according to Figure~\ref{fig:gda}, the average of estimated direct travel times is measured as follows: $\mu = \frac{102 + 84 + 85 + 89}{4} = 90$. The sets and values necessary to determine geographic dispersion are reported in Table~\ref{Table:gpt}.


\begin{table}[H]
\normalsize
\centering
\caption{Sets and values necessary to determine geographic dispersion for instance shown in Figure~\ref{fig:gd}.}
\begin{tabular}{|l|c|c|c|c|}
\hline
Request                         & $N^o_{r}$                                   & $tn^{o}_{r}$             & $N^d_{r}$                     & $tn^{d}_{r}$             \\
\hline
$r_{m}$                         & $\{ \varnothing \}$                          & {0.00}    & $\{ {o_{r_{i}}, o_{r_{k}}} \}$ &  17.00 \\
$r_{i}$ & $\{ {o_{r_{j}}, o_{r_{k}}} \}$               & 10.00  & $\{ {d_{r_{j}}, d_{r_{k}}} \}$ & 13.00  \\
$r_{j}$ & $\{ {o_{r_{i}}, o_{r_{k}}} \}$ & 13.00  & $\{ {d_{r_{i}}, d_{r_{k}}} \}$ & 17.00  \\
$r_{k}$ & $\{ {o_{r_{i}}, d_{r_{m}}} \}$               & 12.00 & $\{ {d_{r_{i}}, d_{r_{j}}} \}$ & 22.00 \\
\hline
\end{tabular}
\label{Table:gpt}
\end{table}

In this table, the first column specifies the request. The second and third columns indicate the $n$ nearest locations to the origin of the request and the average travel time between the origin to these locations, respectively. The fourth and fifth columns report the $n$ nearest locations to the destination of the request and the average travel time between the destination to these locations, respectively. Now, it has become possible to calculate $\omega$, thus: $\omega = \frac{104.00}{8.00} = 13.00$. Finally, the geometric dispersion for this instance is: $gd = 90.00 + 13.00 = 103.00$.




\citet*{reyes2018meal} evaluated the effect of dynamism, urgency and geographic dispersion features on performance metrics towards solutions for the Meal Delivery Routing Problem (MDRP). The MDRP consists of building routes for couriers to deliver meals requested by costumers. Couriers pick up orders as soon as they are available from the restaurant and deliver them at the designated costumer's location. The objective function may include multiple metrics, such as courier compensation, the difference between the announcement of an order and its delivery (click-to-door), and the interval between the release time of an order from the restaurant to arrival time at the drop-off location (ready-to-door). Addressing the MDRP accompanies many challenges, as orders arrive constantly (very dynamic) and are expected to be delivered quickly (high urgency), which poses obstacles to optimize performance measures, specially if the number of couriers is small. The authors observe that higher dynamism increases the route costs, as it is less like to combine orders in a single route to reduce costs. Increases in the geographic dispersion measure are converted in higher click-to-door and ready-to-door mean times. In the meantime, stronger negative effect on these performance measures were observed with increased urgency. 

As previously mentioned, we designed REQreate with the capacity of generating instances with different sizes, levels of urgency, dynamism and geographic dispersion. Consequently, similar studies to \citet{van2016measures} and 
\citet{reyes2018meal} can be performed to explore the potential effects of these measures on solution quality for a comprehensive list of existing and future on-demand transportation problems. A major benefit is to design methods that overcome pitfalls leading to poor performance. For example, upon arrival of very urgent requests it might be interesting to change momentarily the target from building routes with minimal traveled distance to reduce delay times, in favor of meeting a reasonable balance between customer satisfaction and profit for service providers.












\section{Instance similarity}\label{sec:similarity}

To perform informative analyses algorithms should be tested on a diverse benchmark set. The direct rationality is that similar instances are unlikely to contribute with meaningful additional knowledge. Therefore, we present a concept of instance similarity, based on the approach introduced in \citet*{leeftink2018case}. 
In order to assess the similarity of two instances of the same size $I$ and $J$, we take into account that instances for the problems described in this paper consist of a set of requests distributed over a network during a time period. First, we measure the proximity between locations of two distinct requests $r_{i}$ and  $r_{j}$, where $r_{i} \in I$ and $r_{j} \in J$. {We compute and sum the travel times between the origins and destinations of $r_{i}$ and $r_{j}$: $\phi =  tt_{o_{r_{i}}o_{r_{j}}} + tt_{d_{r_{i}}d_{r_{j}}}$.} Requests $r_{i}$ and $r_{j}$ are $\phi$-proximate if $\phi$ is below a given threshold, i.e., $\phi <  th_{tt}$. 
 

On the assumption that two requests are $\phi$-proximate, we also compare if they have a similar time stamp and {earliest departure times}. Let: a) $\tau = \lVert ts_{r_{i}} - ts_{r_{j}} \rVert $; and b) $\vartheta = \lVert e^u_{r_{i}} - e^u_{r_{j}} \rVert$, be the difference of time stamps and {earliest departure times} between $r_{i}$ and $r_{j}$, respectively. Similarly to measuring proximity between location pairs, two requests are $\tau$-proximate and $\vartheta$-proximate if $\tau$ and $\vartheta$ are lower than specified thresholds, i.e., $\tau <  th_{ts}$ and $\vartheta <  th_{e}$. Combining the definitions given so far, the similarity level between two requests $r_{i}$ and  $r_{j}$ ($\xi_{{r_{i}r_{j}}}$) becomes:


\begin{align}
\xi_{{r_{i}r_{j}}} = \begin{cases}
1.00 & \text{if $(\phi < th_{tt})$ $\wedge$ $(\tau <  th_{ts}$) $\wedge$ $(\vartheta <  th_{e})$} \\
0.75 & \text{if $(\phi < th_{tt})$ $\wedge$ ($(\tau <  th_{ts})$ $\oplus$ $(\vartheta <  th_{e})$)} \\
0.50 & \text{if $(\phi < th_{tt})$ $\wedge$ $(\tau \geq th_{ts}$) $\wedge$ $(\vartheta \geq th_{e})$} \\
0.00 & \text{otherwise}
\end{cases}
\end{align}

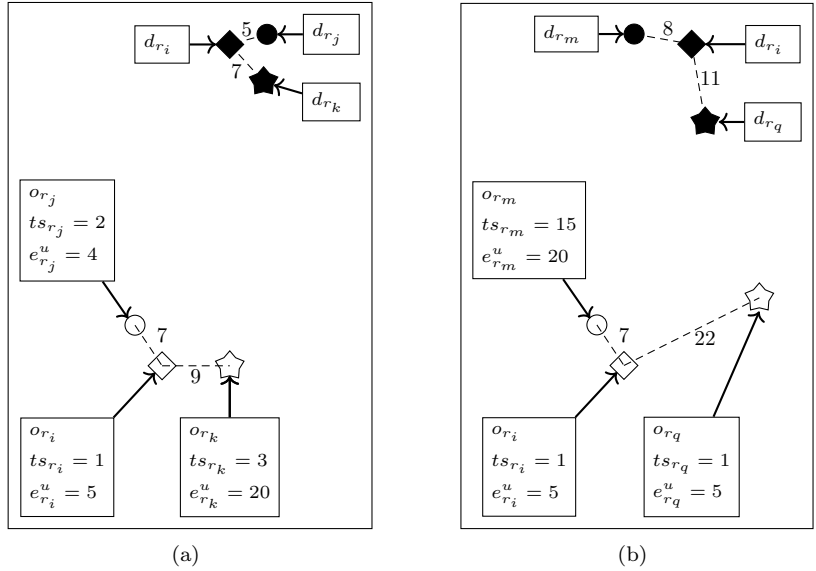
\begin{figure}[H]
\centering
\subfigure[]{

    \begin{tikzpicture}[framed, scale=0.90]
    \node at (1.6, -2.4) [circle, draw, scale=0.8] {};
    \node at (2, -3) [diamond, draw, scale=0.8] {};
    \draw[densely dashed, -] (1.6,-2.4) -- (2, -3);
    \node at (2, -2.55) [scale = 0.8] {7};
    \node at (3, -3)  [star, draw] [scale = 0.8] {};
    \draw[densely dashed, -] (2, -3) -- (3,-3);
    \node at (2.5, -3.15) [scale = 0.8] {9};
    \node at (3.55, 1.9) [circle, draw, scale=0.8, fill=black] {};
    \node at (3, 1.75) [diamond, draw, scale=0.8, fill=black] {};
   \draw[densely dashed, -] (3.55, 1.9) -- (3, 1.75);
    \node at (3.25, 1.98) [scale = 0.8] {5};
    \node at (3.5, 1.2)  [star, draw, fill=black] [scale = 0.8] {};
    \draw[densely dashed, -] (3,1.75) -- (3.5,1.2);
    \node at (3.1, 1.35) [scale = 0.8] {7};
    
     \node[rectangle, draw, align=left] (arr1) at (0.6,-1.0) {\scriptsize 
     $o_{r_{j}}$ \\
     \scriptsize $ts_{r_{j}} = 2$ \\
      \scriptsize $e^u_{r_{j}} = 4$
     };
     \draw[thick,->] (arr1) -- (1.5,-2.3);
      \node[rectangle, draw, align=left] (arr2) at (0.6,-4.5) {\scriptsize 
     $o_{r_{i}}$ \\
     \scriptsize $ts_{r_{i}} = 1$ \\
      \scriptsize $e^u_{r_{i}} = 5$
     };
     \draw[thick,->] (arr2) -- (1.9,-3.1);
      \node[rectangle, draw, align=left] (arr3) at (3,-4.5) {\scriptsize 
     $o_{r_{k}}$ \\
     \scriptsize $ts_{r_{k}} = 3$ \\
      \scriptsize $e^u_{r_{k}} = 20$
     };
     \draw[thick,->] (arr3) -- (3,-3.15);
     \node[rectangle, draw, align=left] (des1) at (4.5,1.9) {\scriptsize 
     $d_{r_{j}}$
     };
     \draw[thick,->] (des1) -- (3.7,1.9);
      \draw[thick,->] (arr3) -- (3,-3.15);
     \node[rectangle, draw, align=left] (des2) at (4.5,0.9) {\scriptsize 
     $d_{r_{k}}$
     };
     \draw[thick,->] (des2) -- (3.65,1.15);
     \node[rectangle, draw, align=left] (des3) at (2.0,1.75) {\scriptsize 
     $d_{r_{i}}$
     };
     \draw[thick,->] (des3) -- (2.8,1.75);
\end{tikzpicture}
\label{fig:sima}
}
 \hspace{0.7cm}
 \subfigure[]{

    \begin{tikzpicture}[framed, scale=0.90]

    \node at (1.6, -2.4) [circle, draw, scale=0.8] {};
    \node at (2, -3) [diamond, draw, scale=0.8] {};
    \draw[densely dashed, -] (1.6,-2.4) -- (2, -3);
    \node at (2, -2.55) [scale = 0.8] {7};
    \node at (4, -2.0)  [star, draw] [scale = 0.8] {};
    \draw[densely dashed, -] (2, -3) -- (4, -2.0);
    \node at (3.2, -2.6) [scale = 0.8] {22};
    \node at (2.15, 1.9) [circle, draw, scale=0.8, fill=black] {};
    \node at (3, 1.75) [diamond, draw, scale=0.8, fill=black] {};
   \draw[densely dashed, -] (2.15, 1.9) -- (3, 1.75);
    \node at (2.65, 1.98) [scale = 0.8] {8};
    \node at (3.2, 0.6)  [star, draw, fill=black] [scale = 0.8] {};
    \draw[densely dashed, -] (3,1.75) -- (3.2,0.6);
    \node at (3.28, 1.25) [scale = 0.8] {11};
    
     \node[rectangle, draw, align=left] (arr1) at (0.6,-1.0) {\scriptsize 
     $o_{r_{m}}$ \\
     \scriptsize $ts_{r_{m}} = 15$ \\
      \scriptsize $e^u_{r_{m}} = 20$
     };
     \draw[thick,->] (arr1) -- (1.5,-2.3);
      \node[rectangle, draw, align=left] (arr2) at (0.6,-4.5) {\scriptsize 
     $o_{r_{i}}$ \\
     \scriptsize $ts_{r_{i}} = 1$ \\
      \scriptsize $e^u_{r_{i}} = 5$
     };
     \draw[thick,->] (arr2) -- (1.9,-3.1);
      \node[rectangle, draw, align=left] (arr3) at (3,-4.5) {\scriptsize 
     $o_{r_{q}}$ \\
     \scriptsize $ts_{r_{q}} = 1$ \\
      \scriptsize $e^u_{r_{q}} = 5$
     };
     \draw[thick,->] (arr3) -- (4.0,-2.2);
     \node[rectangle, draw, align=left] (des1) at (1.15,1.9) {\scriptsize 
     $d_{r_{m}}$
     };
     \draw[thick,->] (des1) -- (2.00,1.9);
     \node[rectangle, draw, align=left] (des2) at (4.2,1.75) {\scriptsize 
     $d_{r_{i}}$
     };
     \draw[thick,->] (des2) -- (3.2,1.75);
     \node[rectangle, draw, align=left] (des3) at (4.2,0.6) {\scriptsize 
     $d_{r_{q}}$
     };
     \draw[thick,->] (des3) -- (3.4,0.6);
\end{tikzpicture}
\label{fig:simb}
}
\caption{Examples of requests with different levels of similarity. }
\label{fig:sim}
\end{figure}

We exemplify the concept of similiarity with Figure~\ref{fig:sim}. In each figure we depict different requests, and we analyze their level of similarity regarding $r_{i}$.  Hypothetically, consider threshold levels to be $th_{tt} = 20$, $th_{ts} = 10$, and $th_{e} = 10$. Identical to Figure~\ref{fig:gd}, travel times between locations are represented by the numbers adjacent to the dashed lines. These values are again measured in fictitious units of time. 

{The values necessary to determine similarity between the pictured requests are reported in Table~\ref{Table:sim}. In this table, the first column specifies which pair of requests the similarity is being computed for. The following three columns give the computed values for $\phi$, $\tau$ and $\vartheta$, respectively. In these columns, the marks \ding{51} and \ding{55} besides the numbers indicate if the values are, respectively, below or above their corresponding thresholds. The last column report the level of similarity between the pair of requests. The first two lines indicate the requests shown in Figure~\ref{fig:sima}. Since all values for requests $r_{i}$ and $r_{j}$ are below the given thresholds, then $\xi_{{r_{i}r_{j}}} = 1.00$. Meanwhile, the difference of earliest departure times ($\vartheta$) is the only value that exceeds the threshold for requests $r_{i}$ and $r_{k}$, therefore $\xi_{{r_{i}r_{k}}} = 0.75$. The final two lines indicate the requests represented in Figure~\ref{fig:simb}. Note that requests $r_{i}$ and $r_{m}$ are only $\phi-$proximate, so $\xi_{{r_{i}r_{m}}} = 0.50$. Meanwhile, the sum distance between locations of $r_{i}$ and $r_{q}$ exceeds $th_{tt}$, for this reason $\xi_{{r_{i}r_{q}}} = 0.00$.}

\begin{table}[H]
\normalsize
\centering
\caption{Computed values necessary to determine similarity between requests shown in Figure~\ref{fig:sim}.}
\begin{tabular}{|l|c|c|c|c|}
\hline
Request pair   & {$\phi$} & {$\tau$} & {$\vartheta$} & {$\xi$} \\
\hline
$r_{i}r_{j}$   & 12.00         \ding{51}              & 1.00  \ding{51}                      & 1.00        \ding{51}                     & 1.00                      \\
$r_{i}r_{k}$ & 16.00   \ding{51}                   & 2.00   \ding{51}                    & 15.00        \ding{55}                    & 0.75                      \\
\hline
$r_{i}r_{m}$ & 15.00         \ding{51}             & 14.00  \ding{55}                    & 15.00   \ding{55}                         & 0.50                      \\
$r_{i}r_{q}$  & 33.00             \ding{55}         & 0.00 \ding{51}                      & 0.00   \ding{51}                          & 0.00         \\
\hline
\end{tabular}
\label{Table:sim}
\end{table}

We introduce the following definition to help determine the similarity between two instances: consider a bipartite graph $G_{s} = (V_{s}, E_{s})$, where the set of vertices $V_{s}$ represent the requests, and there is a non negative weight $w_{e}$ associated with each edge $e \in E_{s}$. The weights represent the similarity level between the two requests of opposite instances. Similarity is computed for every pair of requests between the two instances, and as a result, the same request will most likely exhibit a different level of similarity from multiple requests. Therefore, to determine the similarity between $I$ and $J$, we identify the maximal matching $M_{s} \subset E_{s}$ and measure the average weight of edges: $\Omega_{IJ} = \frac{\sum_{m \in M_s} w_m}{|M|}$.

Essentially, this approach attempts to capture requests that are likely to impose similar restrictions during the execution of the algorithm, e.g., occupy vehicles that travel between the same areas at a similar period of times. We acknowledge the fact that instances sharing the same configuration file (replicas) are expected to have a higher level of similarity when compared to the remainder. Nevertheless, our objective is also to ensure that those instances are still distinct enough to justify performing experiments to extract insightful results. 

\section{Comparison between real data and synthetic instances}\label{sec:comparison}

{In this section, we analyze the mobility patterns of real trips performed by rideshare companies. We use datasets that were made available publicly from \quotes{Chicago, Illinois}~\citep*{chicagodatarideshare}. 
We divide these datasets in different time periods and study statistical properties, such as spatial and distance distributions. We generate synthetic instances that attempt to approximate the observed patterns in the real datasets and discuss some results. The instances for this Section were generated according to attributes for the DARP.} 







 The period selected in which all trips were performed is from September 1, 2019 to September 30, 2019. Ridesharing companies dataset from \quotes{Chicago, Illinois} documents more than 7 million trips during the given period of time. Each record contains information on the pick up and drop off time and locations of the trips, traveled distance, fare paid, along with many other fields. All records are anonymous. Besides, privacy is also preserved by rounding times to the nearest 15 minutes. Therefore, results regarding each of those fields will be an approximation of reality. 




Considering that travel patterns change according to time and day of the week, it is reasonable to conduct this examination on different days and time intervals. First, we split the datasets by days according to working days (business days) and non-working days (weekends and holidays). Second, for working days we extract two time periods of higher demand: a) between 7:00 and 10:00; and b) between 16:00 and 20:00. During non-working days, the time period considered was between 16:00 and 20:00, since in the morning the activity is almost negligible. {To generate the synthetic instances, the origin and destinations were randomly chosen according to the method described in Section~\ref{sec:model}, i.e., following the number of POIs and the best fitted distribution of distances corresponding to each combination of days and time periods.}

\begin{figure}[H]
 \centering
  \subfigure[Spatial distribution between 07:00 and 10:00 during working days.]
 {
\includegraphics[scale =0.30] {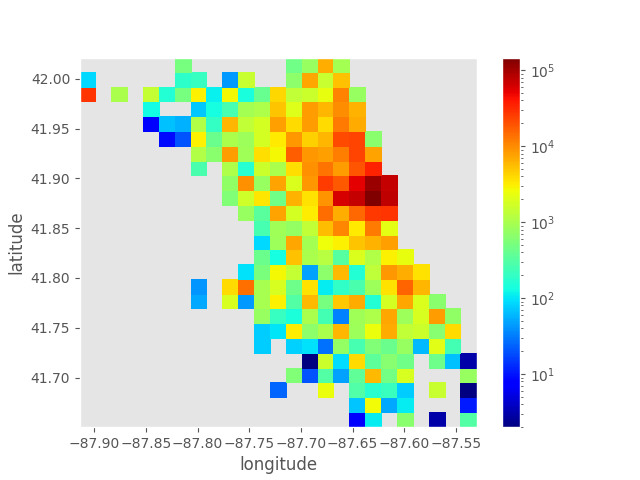}
\includegraphics[scale =0.30] {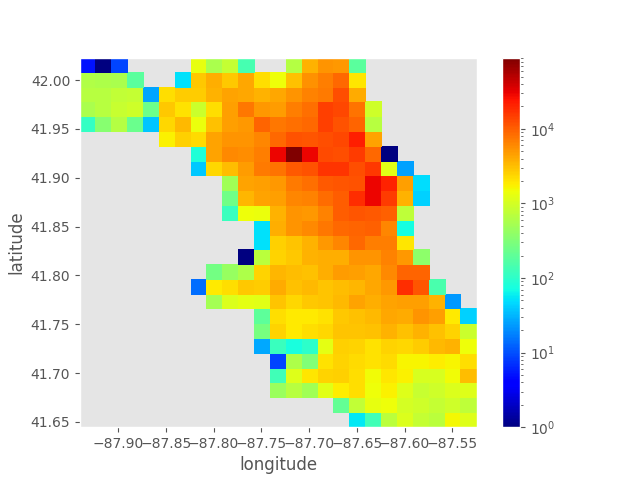}
\label{Fig:heata1620}

}
 \hspace{0.7cm}
 \subfigure[Spatial distribution between 16:00 and 20:00 during working days.]{
 \includegraphics[scale =0.30] {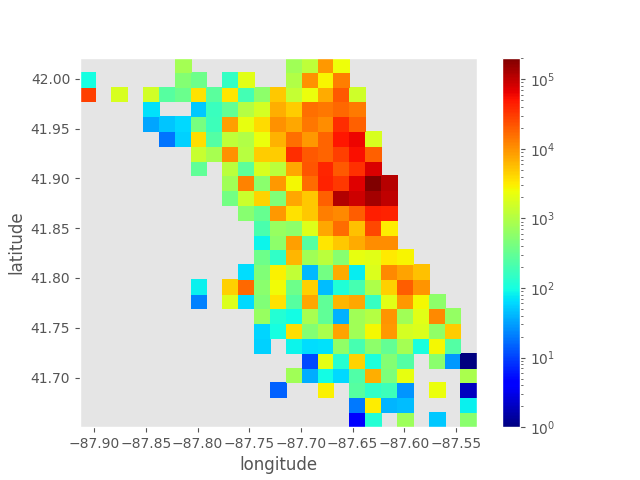}
  \includegraphics[scale =0.30] {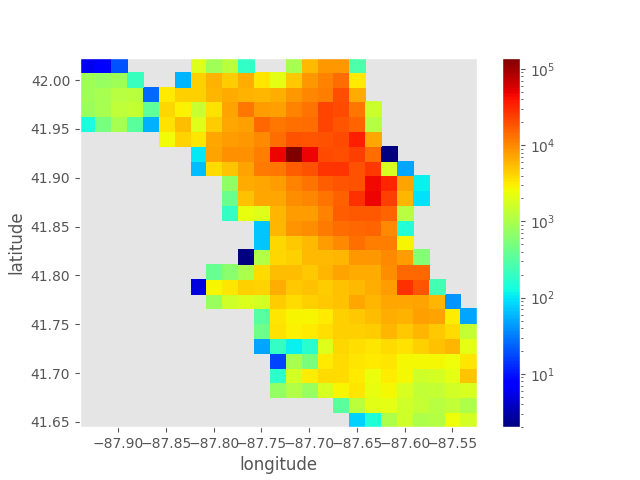}
\label{Fig:heatb1620}
}
 \subfigure[Spatial distribution between 16:00 and 20:00 during non-working days.]{
\includegraphics[scale =0.30] {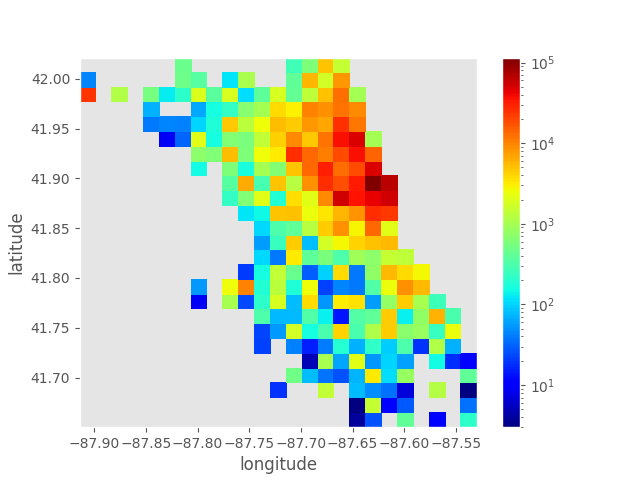}
\includegraphics[scale =0.30] {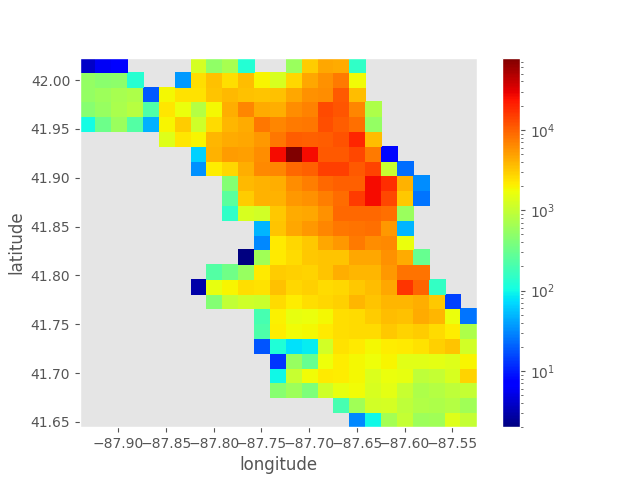}
\label{Fig:heatc1620}
}

\caption{{{Spatial distribution of origin and destination locations. Figures on the left refer to trips reported by rideshare companies in \quotes{Chicago, Illinois}, meanwhile figures on the right concern synthetic generated instances. The color bar accounts for the number of locations.}}}

\label{Fig:heatmap}
\end{figure}

\begin{figure}[H]
 \centering
\includegraphics[scale =0.40] {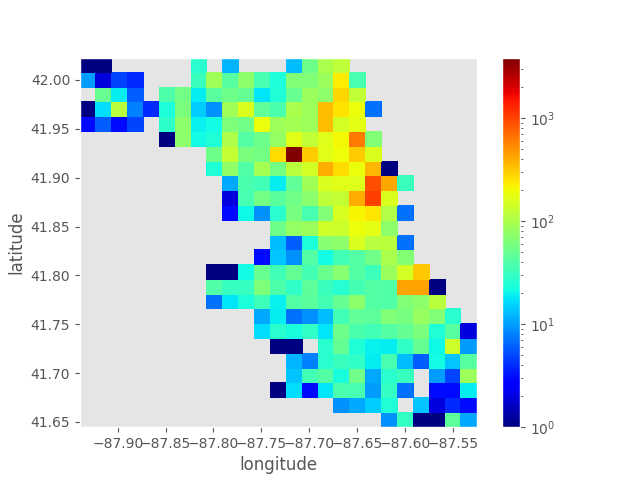}
\caption{{Spatial distribution for Points of Interest (POIs) in the city of \quotes{Chicago, Illinois} according to reported locations in OpenStreetMaps (OSM). The color bar accounts for the number of POIs.}}

\label{Fig:POIsheatmap}
\end{figure}

Figure~\ref{Fig:heatmap} illustrates the spatial distribution of origin and destination locations referring to trips reported by rideshare companies in the city of \quotes{Chicago, Illinois} {and the set of synthetic generated instances.} The three plotted periods for the set of real trips portray similar distribution patterns, as the same regions attracted a large amount of individuals during both working and non-working days (orange and red colored portions). Certainly, this behaviour can be explained as the number of POIs is higher in the aforementioned region (see the yellow area with some orange and red spots in Figure~\ref{Fig:POIsheatmap}). This is further evidence of the positive correlation between spatial distribution of POIs and the potential of a location to be associated with individual's trajectories. Despite being more evenly spread than the locations in the real life datasets, the synthetic instances also display a slightly denser concentration in the region where most of the POIs are situated.

\begin{figure}[H]
 \centering
 \subfigure[]
 {
\includegraphics[scale =0.40] {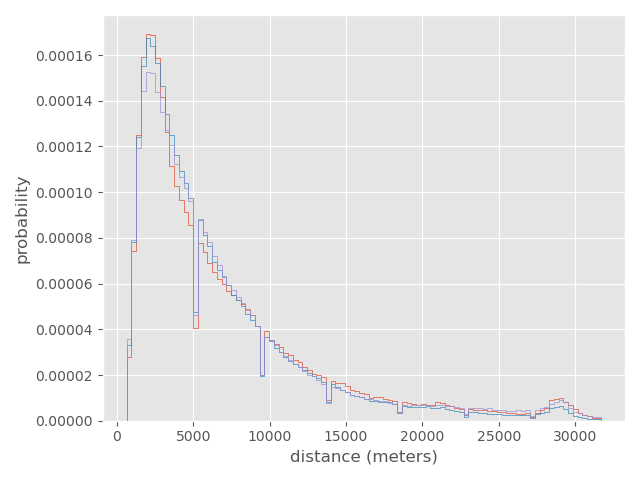}
\label{Fig:disrealx}

}
 \hspace{0.7cm}
 \subfigure[]
 {
 \includegraphics[scale =0.40] {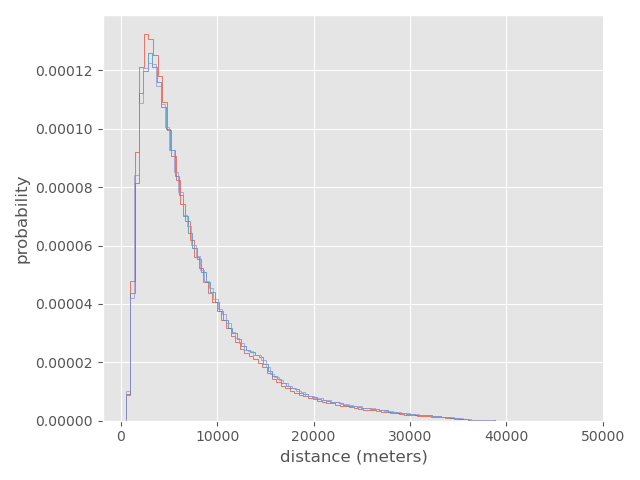}
\label{Fig:distsynx}
}
 {

}

\caption{{Distance distribution referring to trips reported by rideshare companies in the city of \quotes{Chicago, Illinois} (on the left) and for the set of synthetic instances (on the right).}}

\label{Fig:distsyn}
\end{figure}

Distance distributions were also similar during working and non-working days. They are depicted in Figure~\ref{Fig:disrealx}. Such statistical measure capture the extent of the population's interactivity within the urban environment. It is imported to note that the magnitude and shape of the studied area plays an important role, as for example smaller cities impose a more limited upper bound to trip lengths when compared to bigger ones. For the city of \quotes{Chicago, Illinois}, the distances of 75\% of the trips performed by rideshare companies were under 9.17 km for the morning period and below 8.37 km for the afternoon period during working days. Meanwhile, for the interval between 16:00 and 20:00 during non-working days, 75\% of the trips had a smaller length than 8.84 km. A potential reason is that citizens prefer taking part in activities within a short to middle range distance from the areas they spend most of the time (e.g. home and workplace). After reaching its peak (around 2 km), the probability of distance decreases continuously, except when reaching about 27.00 to 30.00 km. This might due to the presence of \quotes{O'Hare International Airport}, as it is located rather far from the city.

As previously indicated, the function and parameters that returns the best fit for the distance distributions are used for the synthetic instances. The distributions used to generate the synthetic instances are illustrated in Figure~\ref{Fig:distsynx}. The shape of the graphs are relatively similar, however, for the synthetic instances, the occurrence of small to medium distances was somewhat lower when compared to the real journeys. 75\% of the distances from generated requests were below 10.22 km and 9.75 km for periods 7:00 to 10:00 and 16:00 to 20:00, respectively, during working days. Also, 75\% of rides for the interval between 16:00 to 20:00 during non-working days reported distances lower than 10.28 km. These values are slightly higher (about 10\% to 15\%) when compared with those of real trips from rideshare companies.

Based on the traveled distances and the duration of the trips, the average speed can be evaluated. Speed values were slightly higher during non-working days when compared to working days. For the latter, the computed averages of speeds were 22.11 km\slash h and 21.30 km\slash h for the intervals between 07:00 to 10:00 and 16:00 to 20:00, respectively. From 16:00 to 20:00 during non-working days, the average speed was 24.79 km\slash h. 

According to Figure~\ref{Fig:POIsheatmap}, one could argue that considering solely the concentration of POIs can possibly improve the spatial distributions of origin and destinations displayed in synthetic instances, however, the distances decay effect presented in Figure~\ref{Fig:distsyn} might be lost. For this reason, new methods that balance both aspects may be elaborated in further research. 

Before concluding this analysis we make three remarks. First, some results are influenced by the method of transport (car, bus, etc.). Since we utilize data from rideshare companies, the trips have also an economic constraint factor, as it is unlikely for citizens to favor this transportation modality as fares rise significantly with long distances. One should bear in mind that an investigation of subway rides, for example, may return a rather different distance distribution, in which higher distances are expected to occur more frequently. {Second, the observed spatial and distance distributions were similar for the studied day and time intervals, however, this might change based on the city.} Finally, the outcome is sensitive to the available information provided concerning POIs locations. Details are provided by OSM users and some of them might be missing. 

\section{{Benchmark set}}

In this section we generate a benchmark set for the ODBRP. Instances are generated according to different sizes, levels of dynamism, urgency and geographic dispersion. The set was built to be diverse so it is possible to evaluate the influence of these properties on proposed methods for the ODBRP. First, we explain how we approximately control the output of each instance property. Then, we describe how instances are organized. 

Size can be easily adjusted with item \textit{requests}, as shown in Section~\ref{sec:generalitems}. Listing~\ref{list:att2} depicts a configuration file for an instance with 500 requests. Meanwhile, dynamism can be regulated by adding a sub-item inside an attribute called \quotes{time\_stamp}, as seen in Listing~\ref{list:att2}. In this example, the instance would have 50\% (0.50) dynamism. We note that for example, an exact 100\% dynamism case can only be generated if the number of dynamic requests is a factor (exact divisor) to the size of the planning period, i.e., exactly one request arriving every $\frac{ts_{max} - ts_{min}}{|R_d|}$ seconds. Therefore one should be careful when setting a specific value for dynamism, since according to the planning period and number of requests, only approximate levels of dynamism are achievable. 

 \begin{lstlisting}[language=json,firstnumber=1, label={list:att2}, caption={Example on how to control properties.}]
{   ...
	"attributes": [
	    "requests": 500,
		{
			"name": "time_stamp",
			"type": "integer",
			"time_unit": "s",
			"pdf": {
                    "type": "uniform",
                    "loc": 30000,
                    "scale": 500
				    },
			"dynamism": 50
		},
		{
			"name": "reaction_time",
			"type": "integer",
			"time_unit": "s",
			"pdf": {
                    "type": "normal",
                    "loc": 600,
                    "scale": 60
				    }
		},
		{
			"name": "latest_departure",
			"type": "integer",
			"time_unit": "s",
			"expression": "time_stamp + reaction_time"
		},
		{
			"name": "direct_travel_time",
			"type": "integer",
			"time_unit": "s",
			"expression": "dtt(origin,destination)",
			"constraints": ["direct_travel_time >= 600", "direct_travel_time <= 1200"]
		}
	],
}
 \end{lstlisting}

The attribute \quotes{reaction\_time} in Listing~\ref{list:att2} controls the levels of urgency. See how \quotes{latest\_departure} is computed as the sum of \quotes{time\_stamp} and \quotes{reaction\_time}, which causes the length of the intervals to perform actions to be equal to attribute \quotes{reaction\_time}. In this example we use a normal distribution (sub-item \textit{pdf}), and the urgency of this instance will have an approximate mean ($\overline{\chi}$) of 10 minutes (600 seconds) and standard deviation (${\chi_s}$) of 1 minute (60 seconds). To obtain higher or lower levels of urgency one must appropriately change these values. Finally, geographic dispersion can be manipulated by declaring an interval of values in the attribute named \quotes{direct\_travel\_time} (see again Listing~\ref{list:att2}), which stores the direct travel time between two locations named \quotes{origin} and \quotes{destination}. The expression includes a predefined function that computes the estimated travel time between the locations. In the given example, every request will have between 10 and 20 minutes of time distance between origin and destination. So, the interval can be adjusted in order to create instances with a greater or smaller value of geographic dispersion.

We used the network from \quotes{Chicago, Illinois} to generate instances. Instances are divided into small (between 300 and 600 requests, in steps of 300), medium (between 900 and 1800 requests, in steps of 300), and large (between 2100 and 3000 requests, in steps of 300). 
The planning period is between 07:00 and 08:00 (1 hour).
The dynamism levels vary between 0\% and 100\% in steps of 10. Meanwhile, urgency mean is set to lie in \{5, 10, 15, 30\} minutes. Meanwhile, the urgency standard deviation is set to lie in \{0, 5, 10\} minutes. Three limited intervals for geographic dispersion were utilized: a) between [180,1000] (short distance trips); b) ]1000,3000] (medium distance trips); and c) ]3000, 6000] (long distance trips). A fourth option with no interval limit is also considered.

Each instance group is identified by $N\_p\_s\_b\_e\_d\_m\_t\_g$, where $N$ denotes name of city for which the network was retrieved; $p$ is an acronym for the problem, in this case ODBRP; $s$ gives the size of the instance; $b$ and $e$ denote the beginning and end of the planning period, respectively; $d$ gives the dynamism level; $m$ and $t$ provide the urgency mean and standard deviation, respectively; and $g$ denotes the geographic dispersion value. We carefully generated the instances so it is possible to fairly evaluate the different levels of dynamism, urgency and geographic dispersion. This means that some instances will portray the same requests locations, and varying only the properties values. 
The instance files are available on Github\footnote{https://github.com/michellqueiroz-ua/instance-generator/tree/master/examples/ODBRP\_benchmark\_configuration\_files}.

\section{Conclusion and future research}\label{sec:finalremarks}

In this paper, we presented REQreate, a tool aimed at generating instances that are significantly more realistic than previous approaches for on-demand public transportation problems. Instances from these problems mainly consist of demand from passengers for transportation. Hence, real life networks are retrieved from OpenStreetMaps (OSM) with support of the OSMnx tool \citep*{boeing2017osmnx}. We described in what manner the attributes and parameters for instances can be given as input to REQreate via a JSON file. Given notations for the Dial-A-Ride Problem (DARP) and On-Demand Bus Routing Problem (ODBRP), we further presented properties that can be considered when analyzing the performance of optimization algorithms: size, dynamism, urgency, and geographic dispersion. The concept of instance similarity was also proposed to provide some level of diversity in the benchmark instance sets.

We used datasets provided by rideshare companies from the city of \quotes{Chicago, Illinois} and examined which statistical properties of human movement could be inferred. Furthermore, REQreate was utilized to generate synthetic instances and we performed a comparison between the properties of the two groups. Requests were {randomly chosen} according to a simple framework taking into account Points of Interest (POIs) and distances as major factors that impact human trajectories. The method has simple and yet reasonable assumptions which can be useful in producing distinct urban patterns when data accessibility is an issue. We also show that the tool can provide constructive suggestions for the planning of on-demand transportation systems. For example, under which spatial, temporal and distance distribution conditions the implementation is profitable. {Finally, we also proposed a benchmark set consisting of 5280 instances for the ODBRP.}

Appropriate future research directions include new methods to describe mathematically a broader variety of human behaviour regarding intra-urban traveling. Hypothetically, population density combined with POIs could strengthen the outcome of spatial distributions. 
Another interesting direction is to understand how the proposed properties (size, dynamism, urgency and geographic dispersion) influence the performance of a newly developed method.


\section*{Acknowledgments}
This project was funded by the University of Antwerp BOF.

\bibliographystyle{apacite}
\bibliography{references}

\appendix
\section*{Appendix}
\addcontentsline{toc}{section}{Appendices}
\renewcommand{\thesubsection}{\Alph{subsection}}
\setcounter{table}{0}
\renewcommand{\thetable}{A\arabic{table}}
\setcounter{figure}{0}
\renewcommand{\thefigure}{B\arabic{figure}}
\setcounter{lstlisting}{0}

\subsection{Configuration file for the DARP} \label{app:A}

The configuration file example for the DARP is divided in Listings~\ref{list:darp1}~and~\ref{list:darp2}. We remark that these Listings are part of the same configuration file, but they are split for visualization purposes. Moreover, this is a simplified version of the configuration file, as items \textit{seed}, \textit{network}, among others were omitted. First, in Listing~\ref{list:darp1}, the planning period is given as input in parameters \quotes{min\_planning\_period} ($ts_{min}$) and \quotes{max\_planning\_period} $ts_{max}$. The declared values will later be used to indicate the interval for request announcements is between 7:00 and 10:00. The vehicles are located in a single depot randomly generated (\quotes{depots}). The \quotes{origin} ($o_{p}$) and \quotes{destination} ($d_{p}$) attributes will be randomly chosen within the boundaries of the network. The integer \quotes{wheelchair\_requirement} attribute will be uniformly chosen for each request inside the interval [0,1], where 1 explicit a requirement for a vehicle that supports a wheelchair, and 0 otherwise. The direct travel time between \quotes{origin} and \quotes{destination} is disclosed in attribute \quotes{direct\_travel\_time}, established by an expression including a predefined function that computes an estimated travel time between two locations: \quotes{dtt(x,y)}, where \quotes{x} and \quotes{y} are attributes or parameters declared with type \quotes{location}. The earliest departure is named \quotes{earliest\_departure} ($e^u_{p}$), the values are {randomly chosen} according to a normal distribution with mean 30600 (8:30) and standard deviation 3600 (1 hour), and the constraint state it must be greater than or equal to parameter \quotes{min\_planning\_period}.

The remainder attributes are declared in Listing~\ref{list:darp2}. The attribute \quotes{time\_stamp} ($ts_{p}$) is specified by the subtraction of \quotes{earliest\_departure} from an attributed named \quotes{lead\_time}  (\quotes{earliest\_departure - lead\_time}), where the latter is uniformly chosen between 0 and 600 seconds (0-10 minutes). Additionally, \quotes{time\_stamp} must also respect the planning period time constraints. Similarly, \quotes{latest\_departure} ($l^u_{p}$) is specified by an expression (\quotes{earliest\_departure + time\_window\_size}), where values for \quotes{time\_window\_size} are randomly chosen within the interval [300,600] seconds (5-10 minutes) according to an uniform distribution. The \quotes{earliest\_arrival} ($e^o_{p}$) is simply calculated from the expression \quotes{earliest\_departure + direct\_travel\_time}. Attribute \quotes{latest\_arrival} ($l^o_{p}$) is calculated by adding \quotes{time\_window\_length} to \quotes{earliest\_arrival}, and is restricted to be  lesser than or equal to parameter \quotes{max\_early\_departure}. The travel time matrix consists of estimated travel times between all pairs of locations in \quotes{depots} ($W$), \quotes{origin} ($O$), and \quotes{destination} ($D$). Finally, we emphasize that each instance can be tested with several combinations of fleet size and vehicle capacity.


 \begin{lstlisting}[language=json,firstnumber=1, label={list:darp1}, float, caption={Configuration file example for the DARP - \textit{part 1}}]
{   ...
	"parameters":[
		{
			"name": "min_planning_period",
			"type": "integer",
			"value": 7,
			"time_unit": "h"
		},
		{
			"name": "max_planning_period",
			"type": "integer",
			"value": 10,
			"time_unit": "h"
		},
		{
			"name": "depots",
			"type": "array_locations", 
			"size": 1,
			"locs": "random"
		}
	],
	"attributes":[
		{  
			"name": "origin",
			"type": "location"
		},
		{  
			"name": "destination",
			"type": "location"
		},
		{
			"name": "wheelchair_requirement",
			"type": "integer",
			"pdf": {
                    "type": "uniform",
                    "loc": 0,
                    "scale": 1
				    }
		},
		{
			"name": "direct_travel_time",
			"type": "integer",
			"time_unit": "s",
			"expression": "dtt(origin,destination)",
			"output_csv": false
		},
		{
			"name": "earliest_departure",
			"type": "integer",
			"time_unit": "s",
			"pdf": {
                    "type": "normal",
                    "loc": 30600,
                    "scale": 3600
				    },
			"constraints": [ "earliest_departure >= min_planning_period"]
		}
		...
	]
    ...
}
 \end{lstlisting}

 \begin{lstlisting}[language=json,firstnumber=1, float, label={list:darp2}, caption={Configuration file example for the DARP - \textit{part 2}}]
{   ...
	"attributes": [
		{
			"name": "lead_time",
			"type": "integer",
			"time_unit": "s",
			"pdf": {
                    "type": "uniform",
                    "loc": 0,
                    "scale": 600
				    },
			"output_csv": false
		},
	   {
			"name": "time_stamp",
			"type": "integer",
			"time_unit": "s",
			"expression": [ "earliest_departure - lead_time"]
			"constraints": [ "time_stamp >= min_planning_period", "time_stamp <= max_planning_period"]
		},
		{
			"name": "time_window_size",
			"type": "integer",
			"time_unit": "s",
			"pdf": {
                    "type": "uniform",
                    "loc": 300,
                    "scale": 300
				    },
			"output_csv": false
		},
		{
			"name": "latest_departure",
			"type": "integer",
			"time_unit": "s",
			"expression": "earliest_departure + time_window_size"
		},
		{
			"name": "earliest_arrival",
			"type": "integer",
			"time_unit": "s",
			"expression": "earliest_departure + direct_travel_time"
		},
		{
			"name": "latest_arrival",
			"type": "integer",
			"time_unit": "s",
			"expression": "earliest_arrival + time_window_size",
			"constraints": [ "latest_arrival <= max_planning_period"]
		}
	],
	"travel_time_matrix": ["depots", "origin", "destination"]
}
 \end{lstlisting}

 \subsection{Configuration file for the ODBRP} \label{app:B}
 
 The configuration file example for the ODBRP is divided in Listings~\ref{list:odbrp1}~and~\ref{list:odbrp2}. As previously remarked in Appendix~\ref{app:A}, these Listings are part of the same configuration file. They were also simplified for visualization purposes. First, in Listing~\ref{list:odbrp1}, a zone named \quotes{zone\_center} is declared in item \textit{places}, where its center point coincides with the network's centroid. The planning period is given as input in parameters \quotes{min\_planning\_period} ($ts_{min}$) and \quotes{max\_planning\_period} $ts_{max}$. The declared values indicate the interval for request announcements is between 6:00 and 9:00. An array of zones named \quotes{zones\_dest} is declared containing only \quotes{zone\_center}. The \quotes{origin} ($o_{p}$) and \quotes{destination} ($d_{p}$) attributes will be randomly chosen within the boundaries of the network, however, \quotes{destination} is bounded to the specific zone declared in array \quotes{zones\_dest}. Suppose this configuration file to be a hypothetical example where the target location of requests is the city center, usually employees commute to  this area during morning peak hours. The direct travel time between \quotes{origin} and \quotes{destination} is disclosed in attribute \quotes{direct\_travel\_time} and was previously described in Appendix~\ref{app:A}. The \quotes{earliest\_departure} ($e^u_{p}$) values are {randomly chosen} according to an uniform distribution within the interval [25200, 32400] (7:00 to 9:00). We remark that constraint \quotes{earliest\_departure $>$= min\_planning\_period} is indeed redundant for this attribute, however we display it in the configuration file for elucidation purposes. 
 
 The remainder attributes are declared in Listing~\ref{list:odbrp2}. First, \quotes{max\_walking} and \quotes{walk\_speed} are default names for attributes that store, correspondingly, the maximum time a passenger is willing to walk to a bus station and their average walking speed. For this example, the values from \quotes{max\_walking} and \quotes{walk\_speed} are {randomly chosen} according to an uniform distributions between 300 to 600 seconds (5-10 minutes) and 4 to 5 km/h, respectively. The set of departure and arrival bus stations are declared in attributes \quotes{stops\_orgn} and \quotes{stops\_dest}, respectively. The expressions that specify both attributes include a predefined function \quotes{stops(x)}, which returns all bus stations within less than \quotes{max\_walking} from location \quotes{x}. The constraints in attributes \quotes{stops\_orgn} and \quotes{stops\_dest} are written with built-in functions of Python\footnote{We strongly recommend getting familiar with the built-in Python functions, as it will help the user grasp the capabilities of the generator.}. The first constraint is \quotes{len(y) $>$ 0}, which guarantees that the array \quotes{y} is not empty. The second constraint \quotes{not (set(y) \& set(z))} guarantees that there is no intersection between the two arrays \quotes{y} and \quotes{z}. The attributes \quotes{time\_stamp} ($ts_{p}$) and \quotes{lead\_time} are declared exactly as in Appendix~\ref{app:A}, except for the addition of sub-item \textit{static\_probability}. By default, \textit{static\_probability} can take the value of a real number within the interval [0,1], expressing the probability of a request being know before the planning period stars (the \quotes{time\_stamp} is set to 0 which represents a static request), therefore allowing to schedule requests in advance. Attribute \quotes{latest\_arrival} ($l^o_{p}$) is calculated by multiplying \quotes{direct\_travel\_time} to 1.5 and adding the result with \quotes{earliest\_departure}. Constraint \quotes{latest\_arrivel $<$= max\_planning\_period} ensures that \quotes{latest\_arrival} does not surpass the planning period time window. The travel time matrix consists of estimated travel times between all pairs in \quotes{bus\_stations}, which is a default name for the bus stations within the boundaries of the network. Finally, we further emphasize that each instance can be tested with several combinations of fleet size and vehicle (bus) capacity.
 
 
 
  \begin{lstlisting}[language=json,firstnumber=1, label={list:odbrp1}, float, caption={Configuration file example for the ODBRP - \textit{part 1}}]
{   ...
    "places": [
		{
			"name": "zone_center",
			"type": "zone",
			"centroid": true,
			"radius": 2000,
			"length_unit": "m"
		}
	],
	"parameters":[
		{
			"name": "min_planning_period",
			"type": "integer",
			"value": 6,
			"time_unit": "h"
		},
		{
			"name": "max_planning_period",
			"type": "integer",
			"value": 9,
			"time_unit": "h"
		},
		{
			"name": "zone_dest",
			"type": "array_zones", 
			"size": 1,
			"value": ["zone_center"]
		},
	],
	"attributes":[
		{  
			"name": "origin",
			"type": "location"
		},
		{  
			"name": "destination",
			"type": "location",
			"subset_zones": "zone_dest",
		},
		{
			"name": "direct_travel_time",
			"type": "integer",
			"time_unit": "s",
			"expression": "dtt(origin,destination)",
			"output_csv": false
		},
		{
			"name": "earliest_departure",
			"type": "integer",
			"time_unit": "s",
			"pdf": {
                    "type": "uniform",
                    "loc": 25200,
                    "scale": 7200
				    },
			"constraints": [ "earliest_departure >= min_planning_period"]
		},
		...
	]
    ...
}
 \end{lstlisting}

 \begin{lstlisting}[language=json,firstnumber=1, float, label={list:odbrp2}, caption={Configuration file example for the ODBRP - \textit{part 2}}]
{   ...
	"attributes": [
		{
			"name": "max_walking",
			"type": "integer",
			"time_unit": "s",
			"pdf": {
    			         "type": "uniform",
        					"loc": 300,
        					"scale": 300
				    },
			"output_csv": false
		},
		{
			"name": "walk_speed",
			"type": "real",
			"time_unit": "kmh",
			"pdf": {
			            "type": "uniform",
        					"loc": 4,
        					"scale": 1
				    }
		},
		{ 
			"name": "stops_orgn",
			"type": "array_primitives",
			"expression": "stops(origin)",
			"constraints": ["len(stops_orgn) > 0"]
		}, 
		{ 
			"name": "stops_dest",
			"type": "array_primitives",
			"expression": "stops(destination)",
			"constraints": ["len(stops_dest) > 0", "not (set(stops_orgn) & set(stops_dest))"]
		}
		{
			"name": "lead_time",
			"type": "integer",
			"time_unit": "s",
			"pdf": {
                    "type": "uniform",
                    "loc": 0,
                    "scale": 600
				    },
			"output_csv": false
		},
	   {
			"name": "time_stamp",
			"type": "integer",
			"time_unit": "s",
			"expression": [ "earliest_departure - lead_time"]
			"constraints": [ "time_stamp >= min_planning_period", "time_stamp <= max_planning_period"],
			"static_probability": 0.5
		},
		{
			"name": "latest_arrival",
			"type": "integer",
			"time_unit": "s",
			"expression": "earliest_departure + (direct_travel_time * 1.5)",
			"constraints": [ "latest_arrival <= max_planning_period"]
		}
	],
	"travel_time_matrix": ["bus_stations"]
}
 \end{lstlisting}

\end{document}